\documentclass[aps,prd,amsfonts,preprint,eqsecnum]{revtex4-1} %
\usepackage{bm}
\usepackage{mathrsfs}
\usepackage{natbib}
\usepackage{graphicx}
\usepackage{amsmath,amssymb,amstext}
\usepackage{enumerate}
\usepackage{subfigure}

\begin{document}
\title{Searching for Gravitational Wave Bursts via Bayesian Nonparametrics Data Analysis with Pulsar Timing Arrays}
\date{\today}

\author{Xihao Deng}
\email{xud104@psu.edu}
\affiliation{Department of Physics, 104 Davey Laboratory, The Pennsylvania State University, University Park, PA  16802-6300, USA}

\begin{abstract}
Gravitational wave burst is a catch-all category for signals whose durations are shorter than the observation period. We apply 
a method new to gravitational wave data analysis --- Bayesian non-parameterics --- to the 
problem of gravitational wave detection, with an emphasis on pulsar timing array observations. In Bayesian non-parametrics,
constraints are set on the function space that may be reasonably thought to characterize the range of gravitational-wave 
signals. This differs from the approaches currently employed or proposed, which focus on introducing parametric signal models 
or looking for excess power as evidence of the presence of a gravitational wave signal. 
Our Bayesian nonparametrics analysis method addresses two issues: (1) investigate if a gravitational wave burst is present in 
the data; (2) infer the sky location of the source and the duration of the burst. 
Compared with the popular method proposed by Finn \& Lommen, our method improves in two aspects: (1) we can estimate the burst duration by adding the prior that the gravitational wave signals are smooth, while Finn \& Lommen ignored this important point; (2) we perform a full Bayesian analysis by marginalizing over all possible parameters and provide robust inference on the presence of gravitational waves, while Finn \& Lommen chose to optimize over parameters, which would increase false alarm risk and also underestimate the parameter uncertainties.  
\end{abstract}



\pacs{}

\maketitle

\section{Introduction}\label{sec:intro}
The Bayesian analysis of gravitational wave detector data for gravitational wave bursts has generally modeled possible bursts in terms of finite set of parameters and their a priori distributions \cite{Seto:2009:sfm, Pshirkov:2010:ogw, cordes:2012:dgw, damour:2001:gwb, siemens:2007:gws, leblond:2009:gwf}. The posterior distribution of the parameters is then determined by the observations. From the posterior parameter distribution either a Bayes Factor or signal-to-noise ratio is estimated and used to decide whether a burst has been detected (e.g.~\cite{finn:2010:dla}). The choice of a finite set of parameters to model the burst imposes unnecessary constraints on the form of the burst as a function of time. Additionally, the Bayes Factor may be ill-defined or overly sensitive to the parameters or their prior distributions (see Chapter 6 in \cite{gelman:2004:bda}). Here we adopt a non-parametric approach to the problem of burst detection: instead of adopting a parameterized model for the burst and identifying parameter priors, we adopt a Gaussian Process prior for the burst function itself and use the observations to constrain the burst as a function of time \cite{rasmussen:2006:gpf}. Gaussian process is defined as a stochastic process for which any finite collection of the samples has a multivariate normal distribution. Besides, as an alternative to the Bayes Factor, we introduce the Deviance Information Criteria (DIC) \cite{spiegelhalter:2002:bmo}, which provides a quantitative measure of how well the data favors the presence of a gravitational wave burst. Finally, we demonstrate our analysis method to simulated pulsar timing array data including a gravitational wave burst. Compared with the popular method proposed by Finn \& Lommen \cite{finn:2010:dla} (denoted as ``F\&L'' in the following), we improve the analysis in two aspects:
\begin{itemize}
\item[(1)] we provide the estimation of the burst durations by incorporating the smoothness of the gravitational wave signals into the prior distribution, while F\&L ignored this important point and resulted in a less informed analysis.
\item[(2)] we marginalize over all parameters in likelihood function and prior distribution when we compute posterior distribution and evidence of gravitational waves. This procedure should be followed to perform a full Bayesian analysis and provide robust inference \cite{gelman:2004:bda}. Nevertheless, F\&L chose to optimize over the parameters in prior distributions, which would increase the false alarm risk and underestimate the parameter uncertainties \cite{rasmussen:2006:gpf}. This is why F\&L obtained strong evidence of gravitational waves and precise estimation of parameters even if the signals are incredibly weak \cite{finn:2010:dla}. 
\end{itemize}
We will discuss these two aspects in details in Section \ref{sec:fl}, and we will demonstrate the advantages of our method over F\&L in Section \ref{sec:demo} by detailed examples. 

Nonparametric data analysis covers techniques that do not assume the model used to fit the data is fixed \cite{gibbons:2003:nsi}. Bayesian nonparametrics stem from seminal work of Ferguson \cite{ferguson:1973:aba} and Doksum \cite{doksum:1974:tan} in early 1970s, which tried to incorporate nonparametric statistics into Bayesian methodology. The advantage of Bayesian nonparametrics over other nonparametric techniques is that it is able to incorporate our expected characteristics of the signals into a prior distribution to set a constraint on the feasible signal patterns we try to search from the data \cite{rasmussen:2006:gpf}. O'Hagan first proposed the general prior distribution for regression analysis, which refers to extract signals from noisy data, is Gaussian process \cite{o'hagan:1978:cfa}. He also proved that the expected characteristics of the signal patterns can be encoded into the mean and covariance of the Gaussian process \cite{o'hagan:1978:cfa}. This approach has then been applied to solve various regression problems in geostatistics, meteorology, computer science, machine learning, etc. Here we apply Bayesian nonparametric regression to search for gravitational wave bursts from pulsar timing array data.

In Section \ref{sec:bayes}, we describe the general principles and methodology of Bayesian nonparametric analysis. In Section \ref{sec:app} we apply this method to analyze simulated pulsar timing array data including the contribution from a gravitational wave burst. In Section \ref{sec:demo} we illustrate the effectiveness of this analysis by applying it to several representative examples. We will also demonstrate the strength of our method over F\&L. Finally, we summarize our conclusion in Section \ref{sec:concl}.



\section{Bayesian Nonparametric Methodology} \label{sec:bayes}
Bayesian nonparametric analysis is introduced to analyze the data when analysts cannot model the data by a set of fixed number of parameters \cite{hjort:2010:bn}. Here we give a brief description of the basic framework of this methodology, applied to the analysis of time series data that may include a signal we are trying to detect. We refer readers to \cite{ghosh:2003:bn, rasmussen:2006:gpf, hjort:2010:bn} for details and other applications of Bayesian nonparametrics.

\subsection{Framework of Bayesian Nonparametric Analysis} \label{sec:framework}
When we cannot model the signal by a fixed number of parameters, Bayesian nonparametric approach assigns a prior distribution on the signal itself and then infer its pattern \cite{ghosh:2003:bn, rasmussen:2006:gpf}. For example, when time series data $\bm{\mathrm{y}}$ contains a signal denoted by a vector $\bm{f}$ and additive zero mean noises, we can assign a prior distribution $q(\bm{f})$ on $\bm{f}$, and according to Bayes' theorem, the posterior probability density of $\bm{f}$ would be:
\begin{equation} \label{eq:posterior0}
p(\bm{f}|\bm{\mathrm{y}}) \propto \mathrm{\Lambda}(\bm{\mathrm{y}}|\bm{f})q(\bm{f})
\end{equation}
where $\Lambda(\bm{\mathrm{y}}|\bm{f})$ is the likelihood function, which will be a Gaussian distribution of $\bm{\mathrm{y}}$ if the noises are Gaussian distributed. Bayesian nonparametric inference will first choose a prior $q$ based on our expectation of $\bm{f}$, and then infer $\bm{f}$ by computing posterior $p$.



\subsubsection{Prior Probability Density $q$} \label{sec:prior}
The probability density $q(\bm{f})$ describes our expectations of the signal before we analyze the data set $\bm{\mathrm{y}}$. It plays the key role in Bayesian nonparametric analysis since it would set a constraint on the feasible function forms we try to explore \cite{sudderth:2006:gmf}. 

In general, we may write $\bm{f}$ in a discrete Fourier transform since the observation times are discrete \cite{summerscales:2008:mef,Bretthorst:1988:bsa},
\begin{equation}\label{eq:fourier1}
 f_i = \sum_{k}^N A_k\cos(\omega_k t_i) + B_k\sin(\omega_k t_i) 
\end{equation}
for signal $\bm{f}$ beginning at time $t_0$ and ending at some later time $t_0+T$, and with $t_i$ the observation times in the interval $[t_0,t_0+T]$. Under the minimal assumption that there is no preferred signal starting time $t_0$ or  duration it is straightforward to find an ``ignorance prior'' for each of the coefficients $A_k$ and $B_k$ \cite{Bretthorst:1988:bsa, summerscales:2008:mef}:
\begin{align}
Q_k(A_k|\sigma_k) &= \frac{\exp(-A^2_k/2\sigma^2_k)}{\sqrt{2\pi}\sigma_k} \nonumber \\
Q_k(B_k|\sigma_k) &= \frac{\exp(-B^2_k/2\sigma^2_k)}{\sqrt{2\pi}\sigma_k}
\end{align}
with $N$ and $\sigma_k$ undetermined. 

In conventional regression problem we would fix $N$ and choose some prior for the $\sigma_k$ \cite{Bretthorst:1988:bsa}. Instead, however, let us take a different approach. Noting that the $A_k$ and $B_k$ are Gaussian random variables we may regard $\bm{f}$ as a Gaussian process with correlation function
\begin{subequations}
\begin{align} \label{eq:kernelg}
 \mathrm{K}_{lm} &= \langle f_l\,f_m \rangle \nonumber \\
&= \sum_k \sigma^2_{k}\cos(\omega_k t_l)\cos(\omega_k t_m) + \sigma^2_{k}\sin(\omega_k t_l)\sin(\omega_k t_m) \nonumber \\
                 &= \sum_k \sigma^2_{k}\cos\left[\omega_k(t_l-t_m)\right]
\end{align}
The covariance $\mathrm{K}_{lm}$ is referred to as the kernel of the Gaussian process prior and it has to be positive semidefinite  \cite{rasmussen:2006:gpf}. Correspondingly, the prior of $\bm{f}$ can be written as
\begin{align} \label{eq:priorf}
 q(\bm{f}|\bm{\theta}) &= N(\bm{f}|\bm{\mathrm{K}}) \nonumber \\
&= \frac{\exp\left(-\frac{1}{2}\bm{f}^{T}\bm{\mathrm{K}}^{-1}\bm{f}\right)}{\sqrt{(2\pi)^{\mathrm{dim}\,\bm{\mathrm{f}}}\,\det||\bm{\mathrm{K}}||}} \end{align}
\end{subequations}
where $\bm{\theta}$ denotes the unknown parameters embedded in the kernel such as $\sigma_k$. Such parameters are referred to as {\it hyperparameters} \cite{gelman:2004:bda}. In full Bayesian inference, we also need to choose a prior probability density $q_\theta$, i.e., {\it hyperprior}, for the hyperparameter set \cite{rasmussen:2006:gpf, gelman:2004:bda}, and the joint prior of $\bm{f}$ and the hyperparameters would be
\begin{equation} \label{eq:jprior}
 q_0(\bm{f},\bm{\theta}) = q(\bm{f}|\bm{\theta}) q_\theta(\bm{\theta})
\end{equation}
Correspondingly we will infer hyperparameters together with $\bm{f}$. We can include our expectation of the signal in the prior probability density $q$ by choosing a specific kernel and a specific hyperprior.

To recap, our non-parametric analysis has characterized the signal by the parameter set $\bm{f}$ and set its prior distribution as a Gaussian process. Unlike conventional regression approaches there is no predetermined mathematical form of $\bm{f}$ aside from those we impose via a prior on the kernel $\bm{\mathrm{K}}$. An appropriate choice of $\bm{\mathrm{K}}$ can set a strong constraint on the smoothness, the trend and the variations of the signal patterns \cite{rasmussen:2006:gpf}.


\subsubsection{Bayesian Nonparametric Inference} \label{sec:inference0}
Since we have chosen the appropriate priors, we can make the inference of $\bm{f}$, i.e., the function form of the signal. Assuming the noises are Gaussian distributed with zero mean, the joint posterior probability density of $\bm{f}$ and hyperparameters would be
\begin{subequations}\label{eq:inference0}
\begin{align} \label{eq:poster0}
p(\bm{f},\bm{\theta}|\bm{\mathrm{y}}) &= \frac{1}{Z(\bm{\mathrm{y}})} \Lambda(\bm{\mathrm{y}}|\bm{f})q(\bm{f}|\bm{\theta})q_\theta(\bm{\theta}) \nonumber \\
&= \sqrt{\frac{\det||\bm{\mathrm{A}}||}{(2\pi)^{\mathrm{dim}\bm{\mathrm{A}}}}}\exp\left[-\frac{1}{2}(\bm{f}-\bm{f}_{m})^{T}\bm{\mathrm{A}}(\bm{f}-\bm{f}_{m})\right] \nonumber \\
&\times \frac{1}{Z(\bm{\mathrm{y}})}\Lambda_{\theta}(\bm{\mathrm{y}}|\bm{\theta})q_\theta(\bm{\theta})
\end{align}
where $Z(\bm{\mathrm{y}})$ is the normalization constant; $\bm{\mathrm{A}}$ is 
\begin{equation}
\bm{\mathrm{A}} = \bm{\mathrm{K}}^{-1} + \bm{\mathrm{C}}^{-1}
\end{equation}
with $\bm{\mathrm{C}}$ denoting the noise covariance matrix; and $\bm{f}_m$ satisfies
\begin{equation}
\bm{\mathrm{A}} \bm{f}_m = \bm{\mathrm{C}}^{-1} \bm{\mathrm{y}}
\end{equation}
$\Lambda_{\theta}(\bm{\mathrm{y}}|\bm{\theta})$ is the likelihood function of hyperparameters after marginalizing over $\bm{f}$
\begin{align}
\Lambda_{\theta}(\bm{\mathrm{y}}|\bm{\theta}) &= \int \Lambda(\bm{\mathrm{y}}|\bm{f})q(\bm{f}|\bm{\theta}) \, \mathrm{d}\bm{f} \nonumber \\
&= \frac{\exp\left[-\frac{1}{2}\bm{\mathrm{y}}^{T}\bm{\mathrm{C}}^{-1}\bm{\mathrm{y}}\right]}{\sqrt{(2\pi)^{\mathrm{dim}\,\bm{\mathrm{y}}}\det||\bm{\mathrm{C}}||}} \times \frac{\exp\left[\frac{1}{2}(\bm{\mathrm{C}}^{-1}\bm{\mathrm{y}})^{T}\bm{\mathrm{A}}^{-1}(\bm{\mathrm{C}}^{-1}\bm{\mathrm{y}})\right]}{\sqrt{\det||\bm{\mathrm{A}}||\det||\bm{\mathrm{K}}||}}
\end{align}
If we would like to infer $\bm{f}$, we need to choose a hyperprior $q_\theta$ and marginalize over hyperparameter $\bm{\theta}$ to obtain marginalized posterior for $\bm{f}$, i.e., 
\begin{equation}
p_\theta(\bm{f}|\bm{\mathrm{y}}) = \int p(\bm{f},\bm{\theta}|\bm{\mathrm{y}}) \, \mathrm{d}\bm{\theta} 
\end{equation}
and if we are also interested in estimating hyperparameter $\bm{\theta}$, we need to marginalize over $\bm{f}$ and obtain the posterior for $\bm{\theta}$, i.e., 
\begin{equation}
p_\theta(\bm{\theta}|\bm{\mathrm{y}}) = \int p(\bm{f},\bm{\theta}|\bm{\mathrm{y}}) \, \mathrm{d}\bm{f} = \frac{1}{Z(\bm{\mathrm{y}})}\Lambda_{\theta}(\bm{\mathrm{y}}|\bm{\theta})q_\theta(\bm{\theta})
\end{equation}
\end{subequations}
Eq.~\eqref{eq:inference0} summarizes Bayesian nonparametric inference, which gives estimation on the signal $\bm{f}$ and hyperparameters.

\subsection{Comparison with Bayesian Parametric Analysis} \label{sec:comparison}
At this point, it is worth comparing the Bayesian nonparametric inference described above and the conventional inference methods that try to fit the data by a model with a fixed number of parameters. 

The conventional inference methods assume that we know the analytical formulae of the signals we try to detect and we can characterize them with a fixed number of parameters. For example, to detect gravitational waves from a non-spinning binary black hole inspiral, we can characterize the signal by post-Newtonian formula with 8 parameters to characterize the signal --- chirp mass, mass ratio, coalescence time, strain amplitude, sky location of the source, polarization and inclination angle of the orbital plane \cite{blanchet:2006:grf}. In this way, we can obtain key characteristics about the signals since the parameters we use to model the signal usually have explicit physical meanings. However, this approach requires that we are able to achieve the analytical formulae of the signals. 

When the analytical formulae of the signals are not available, the conventional approach could model the signal by a linear superposition of a finite number of basis functions (e.g. Chapter 16 of \cite{gelman:2004:bda}), i.e.,
\begin{equation}
 f_i = \sum^{N}_k \alpha_k \Phi_k(t_i) 
\end{equation}
where $\Phi_k$ is the basis function and $\alpha_k$ is the corresponding coefficient that would be the unknown parameters we try to infer; $N$ is the number of the basis functions, which has to be fixed. For example, we can choose $\Phi_k$ as the Fourier modes like Eq.~\eqref{eq:fourier1}. However, if we do not know much information of the signals, we do not know how many basis functions we need to choose. A model with too many basis functions would be so complex as to overfit the data while a model with too few basis functions would be so simple as to underfit the data.

In contrast, Baysian nonparametric approach directly infers the signal pattern by
assigning Gaussian process prior distributions to set strong constraints on it, which would avoid using a detailed physical model. However, since Bayesian nonparametrics do not use the parameters that describe the physical characteristics of the signal like conventional methods, it would provide less direct insight for the signal.

The advantages and disadvantages of Bayesian nonparametric analysis and conventional approach are summarized in the following:
\begin{itemize}
\item[(1)] Bayesian nonparametric analysis is more effective to infer the signals when we do not know their analytical formulae. However, it provides less physical characteristics of the signals than the conventional approach.
\item[(2)] The conventional approach would infer the physical parameters that characterize the signals, which provides more direct insight than Bayesian nonparametric approach. However, it requires that we have adequate information to model the signals with a fixed number of physical parameters, which is not always available. 
\end{itemize}



\section{Bayesian Nonparametric Analysis on Gravitational Wave Bursts with Pulsar Timing Arrays} \label{sec:app}
Pulsar timing data collected from an individual pulsar consists of a time series of pulse time of arrival (TOA) measurements, which 
are compared with the predicted arrival times based on a timing model including all non-gravitational-wave effects. The difference  
between the observed and expected pulse arrival times are referred to as timing residuals. A collection of the timing residuals obtained 
from an array of pulsars would include timing noises which are uncorrelated among different pulsars, and potentially gravitational 
wave effects which are correlated among the pulsar timing array. To seek the evidence of a gravitational wave burst, we need to match 
the timing residuals with a model that characterizes the contribution of the burst. However in general cases, we do not have physical models for the burst sources and so we are not able to characterize the gravitational wave burst by some analytical formula or by a fixed number of basis functions. Correspondingly, as discussed in Sec.~\ref{sec:comparison}, the conventional approach is not applicable. Therefore, 
we introduce Bayesian nonparametric analysis described in the last section to detect and characterize gravitational wave bursts with a 
pulsar timing array.

\subsection{Properties of the Pulsar Timing Response to the Passage of A Gravitational Wave Burst}
A plane gravitational wave propagating in direction $\hat{k}$ is represented by the transverse-traceless gauge metric perturbation \cite{misner:1973:g}
\begin{align}
\mathbf{h}_{lm}(t,\vec{x}) &= 
h_{(+)}(t-\hat{k}\cdot\vec{x}) e^{(+)}_{lm} +
h_{(\times)}(t-\hat{k}\cdot\vec{x}) e^{(\times)}_{lm}
\end{align}
where $e^{(A)}_{lm}$ is the polarization tensor. Following \cite{finn:2010:dla}, the $j$th pulsar timing response to such a plane gravitational wave can be written as
\begin{subequations}
\begin{equation}\label{eq:residual}
\tau_{j}(t)
= -\frac{1}{2}\hat{n}_{j}^l\hat{n}_{j}^m\big[e^{(+)}_{lm}\mathcal{H}_{j(+)}+e^{(\times)}_{lm}\mathcal{H}_{j(\times)} \big] 
\end{equation}
where $\hat{n}_j$ the direction from Earth toward the $j$th pulsar and $\mathcal{H}_{(A)}$ is 
\begin{align} \label{eq:H2}
\mathcal{H}_{j(A)}(t) &= \frac{\tau_{(A)}(t)}{1+\hat{k}\cdot\hat{n}_j} - \frac{\tau_{(A)}(
t-L(1+\hat{k}\cdot\hat{n}_j))}{1+\hat{k}\cdot\hat{n}_j}
\end{align}
where $\tau_{(A)}$ is the integral of $h_{(A)}$,
\begin{equation} \label{eq:tauP}
\frac{\mathrm{d}\,\tau_{(A)}}{\mathrm{d}\,u} = h_{(A)}(u) 
\end{equation}
\end{subequations}
We can see that the gravitational wave contribution is the sum of two functionally identical terms, one time-shifted with respect to the other by an amount proportional to the Earth-pulsar distance along the wave propagation direction. The first term is referred to as the ``Earth Term'', while the second is referred to as the ``Pulsar Term.''

When the duration of the gravitational wave bursts $\Delta T$ and the observational duration of pulsar timing array data $T$ are much less than $L(1+\hat{k}\cdot\hat{n})$, it is most likely that only the Earth term contributes to the correlated timing residuals \cite{finn:2010:dla}, unless (1) there are fortuitous lines of sight of a pair or more of the pulsars where the time delay among 
those Pulsar terms is small enough to be within the observational duration \cite{pitkin:2012:egw}; (2) there are a large number of pulsars providing a long time baseline that might be able to detect more burst sources \cite{cordes:2012:dgw}. However, in the current international pulsar timing array, there are not so many pulsars whose timing noises are low enough to be effective in gravitational wave detection \cite{hobbs:2010:tip, demorest:2012:lot, manchester:2012:tpp}, and it is uncertain when we can have a pulsar timing array with a large number of low timing noise pulsars. Therefore, in this paper, we only consider the effects of the Earth term on the detection of gravitational wave bursts, and the corresponding pulsar timing response can be written as

\begin{subequations}\label{eq:tau}
\begin{equation}
\tau_{j}(t) = F_j^{(+)}\tau_{(+)}(t) + F_j^{(\times)}\tau_{(\times)}(t)
\end{equation}
where $F_j^{(A)}$ is the pattern function of the $j$th pulsar,
\begin{equation}
F_j^{(A)} = -\frac{\hat{n}_j^l\hat{n}_j^m e^{(A)}_{lm}}{2(1+\hat{k}\cdot\hat{n}_j)}
\end{equation}
\end{subequations}

In the pulsar timing array waveband, the internal motion of the gravitational wave sources is expected to be smooth with 
the evolution of time \cite{hobbs:2010:tip}. Correspondingly, the waveform $h_{(A)}(t)$ of a gravitational wave burst is 
expected to be a smooth funtion of the observation times \cite{misner:1973:g} and thus $\tau_{(A)}(t)$ should also be a smooth function of time. 

Furthermore, for a detectable gravitational wave burst through pulsar timing arrays, the duration of the burst should be 
shorter than observation duration, or otherwise it is not considered as ``burst'' signal. 

These two properties of the timing residual are very important and they are considered as our prior knowledge of the gravitational wave bursts. As we will see in the next subsection, we will take advantage of this prior information to model the pulsar timing response to the passage of a gravitational wave burst.

\subsection{The Choice of Prior Probability Distribution} \label{sec:priorgw}
\subsubsection{Priors of $\tau_{(+)}$ and $\tau_{(\times)}$} \label{sec:priortau0} 
As described in Sec.~\ref{sec:bayes}, to apply Bayesian nonparametrics to detect gravitational wave bursts, we need to choose an appropriate prior probability to constrain feasible function forms of the signals. For $\tau_{(+)}$ and $\tau_{(\times)}$, we have 3 expectations as discussed in the last subsection
\begin{itemize} \label{it:char}
\item[(1)] They are the same for different pulsars.
\item[(2)] They are smooth functions of time.
\item[(3)] They have a characteristic time duration that is shorter than the observation time.
\end{itemize}
We need to choose their Gaussian process priors with appropriate kernels to fulfill these three expectations. 

We do not have any information on the arrival time of the gravitational wave burst, which means that priors of $\tau_{(+)}$ and 
$\tau_{(\times)}$ should hold the time translational symmetry \cite{summerscales:2008:mef} and the kernels should only depend on 
the difference between observation times, i.e., stationary \cite{rasmussen:2006:gpf}. $\tau_{(+)}$ and $\tau_{(\times)}$ are also 
expected to be infinitely differentiable on time; correspondingly, their mean squares under their Gaussian process priors have to 
be infinitely differentiable, which requires their stationary kernels $\mathrm{K}_{(A)}(\Delta t)$ infinitely differentiable at 
$\Delta t = 0$  \cite{adler:1981:tgo}, where $\Delta t$ denotes difference between any two observation times. This is a very 
rigorous requirement since the stationary kernels have to be both positive semidefinite and infinitely differentiable, and only 
few of kernels we know satisfy it
\cite{stein:1999:ios}. The one with the least number of hyperparameters is the square exponential kernel \cite{rasmussen:2006:gpf}
\begin{align} \label{eq:kernelW}
\mathrm{K}_{(+)}(\Delta t) &= \sigma^{2}_{+}\exp(-\frac{\Delta t^2}{2\,\lambda^2}) \\
\mathrm{K}_{(\times)}(\Delta t) &= \sigma^{2}_{\times}\exp(-\frac{\Delta t^2}{2\,\lambda^2})
\end{align}
where $\sigma_{+}$, $\sigma_{\times}$ and $\lambda$ are hyperparameters. We assign two different kernels to two polarization components of 
the gravitational wave burst because the two components are independent of each other \cite{misner:1973:g}. Correspondingly, the two rms amplitudes $\sigma_{+}$ and $\sigma_\times$ are two independent hyperparameters. The two polarizations can be transformed into each other by basis rotation in the sky plane \cite{misner:1973:g} so one might think that we can set $\sigma_{+}=\sigma_\times$ by fixing a rotation angle. However, in this case, we have to treat the rotation angle (polarization angle) as another independent hyperparameter, which indicates that there have to be two independent hyperparameters to characterize the amplitudes of the two polarization components. The hyperparameter $\lambda$ is the characteristic time scale of the burst, 
which characterizes temporal correlation of $\tau_{(+)}$ and $\tau_{(\times)}$ along the observation times. When $\Delta t >\sqrt{2}\lambda$, 
the values of the kernels would exponentially damp, and $\tau_{(A)}(t)$ and $\tau_{(A)}(t+\Delta t)$ would be almost 
uncorrelated. Correspondingly, $\sqrt{2}\lambda$ should be approximately the duration of the burst and $\lambda$ should be shorter than 
the observation duration. Here we assume that $\lambda$ is the same for both of the two polarization components because the duration of 
the burst scales with the ratio of the typical radius to the internal velocity of the source \cite{finn:2010:dla}, which is the same for the two components. 

We can see the kernels Eq.~\eqref{eq:kernelW} are special cases of the general form Eq.~\eqref{eq:kernelg} by performing a discrete Fourier transform
\begin{subequations}
\begin{equation}
\mathrm{K}_{(+,\times)}(\Delta t) = \sum_k \sigma^2_{+,\times} \exp\left(-\frac{\lambda^2\omega^2_k}{2}\right) \cos(\omega_k \Delta t)
\end{equation}
compared with Eq.~\eqref{eq:kernelg}, choosing the special kernels as Eq.~\eqref{eq:kernelW} is equivalent to setting $\sigma^2_k$ in Eq.~\eqref{eq:kernelg} as
\begin{equation}
\sigma^2_k = \sigma^2_{+,\times} \exp\left(-\frac{\lambda^2\omega^2_k}{2}\right)
\end{equation} 
\end{subequations}

To recap, the priors of $\tau_{(+)}$ and $\tau_{(\times)}$ would be two zero mean Gaussian processes with kernels in Eq.~\eqref{eq:kernelW}. They fulfill the three expectations stated above as
\begin{itemize}
\item[(1)] the priors of $\tau_{(+,\times)}$ are independent of pulsars.
\item[(2)] kernels Eq.~\eqref{eq:kernelW} guarantee that $\tau_{(+,\times)}$ sampled from the Gaussian process priors are most likely smooth functions of times since the kernels are infinitely differentiable at $\Delta t = 0$.
\item[(3)] the hyperparameter $\lambda$ in kernels Eq.~\eqref{eq:kernelW} characterizes the characteristic time-scale of the burst.
\end{itemize}

 \subsubsection{Priors of Timing Residuals Induced by A Gravitational Wave Burst} 

Now we need to choose an appropriate prior probability density for the pulsar timing residuals induced by a gravitational wave burst $\bm{\tau}$, i.e., the prior distribution of the signal. Because $\bm{\tau}$ is the linear superposition
of $\tau_{(+)}$ and $\tau_{(\times)}$ (see Eq.~\eqref{eq:tau}), whose priors are independent zero mean Gaussian process priors, so 
the prior of $\bm{\tau}$ should also be a zero mean Gaussian process prior, and its kernel should be the linear superposition 
of the two kernels of the Gaussian process priors of $\tau_{(+)}$ and $\tau_{(\times)}$, i.e., 
\begin{subequations} \label{eq:priortau}
\begin{equation}
q(\bm{\tau}|\hat{k}) = \frac{\exp\left[-\frac{1}{2}\bm{\tau}^{T}\bm{\mathrm{K}}^{-1}\bm{\tau}\right]}{\sqrt{(2\pi)^{\mathrm{dim}\bm{\mathrm{K}}}\det||\bm{\mathrm{K}}||}}
\end{equation}
where $\bm{\mathrm{K}}$ is expressed as
\begin{equation} \label{eq:kernel}
\mathrm{K}_{j(\alpha),\,k(\beta)} = \left(\sigma^{2}_{+}F^{(+)}_j F^{(+)}_k + \sigma^{2}_{\times}F^{(\times)}_j F^{(\times)}_k\right) \exp\left[-\frac{(t_{j(\alpha)}-t_{k(\beta)})^2}{2\,\lambda^2}\right]
\end{equation}
where $j,\, k$ are pulsar indices and $\alpha,\,\beta$ are the indices for the observation times of pulsar timing measurements. Here we have
used Eq.~\eqref{eq:tau} and Eq.~\eqref{eq:kernelW}.
\end{subequations}

\subsubsection{Prior of Hyperparameters} \label{sec:hyperprior}
The prior probability density of $\bm{\tau}$, i.e., Eq.~\eqref{eq:priortau}, contains several hyperparameters. We also need to choose appropriate prior probability densities for those hyperparameters. 
According to Eq.~\eqref{eq:priortau}, we have 5 hyperparameters --- rms gravitational wave amplitudes $\sigma_{+}$ and 
$\sigma_{\times}$, characterisitic time-scale $\lambda$, and gravitational wave propagation direction $\hat{k}$ which is 
embedded in the pattern function $F^{(+)}$ and $F^{(\times)}$. 



We first choose the hyperpriors for $\sigma_{+}$ and $\sigma_{\times}$. The pulsar timing array data is assumed to be normally distributed with the mean $\bm{\tau}$ due to the assumption that the noises are normally distributed, and the prior distribution of $\bm{\tau}$ is also chosen to be a Gaussian process. This is a two-level normal model and the choice of hyperprior on the rms amplitude is discussed in detail in \cite{gelman:2006:pdf}. Both $\sigma_{+}$ and $\sigma_{\times}$ appear like scale parameters, suggesting the use of the Jeffreys prior \cite{jeffreys:1946:aif}; however, the Jeffreys prior leads to a non-normalizable posterior probability density \cite{gelman:2006:pdf}, which means it is not an appropriate choice for hyperpriors of $\sigma_{+}$ and $\sigma_{\times}$. If the dimension of $\bm{\tau}$, which in our case equals the number of data points, is larger than 5, a uniform hyperprior distribution of the rms amplitudes is recommended \cite{gelman:2006:pdf}, i.e., 
\begin{subequations} \label{eq:hyperprior}
\begin{equation} \label{eq:hyperpriors}
q_{+}(\sigma_{+}), q_{\times}(\sigma_{\times}) \propto 1
\end{equation}

We then choose the hyperprior for the characteristic time-scale $\lambda$. We expect $\lambda$ should be shorter than the observation duration $T$. As discussed before, $\sqrt{2}\lambda$ characterizes the duration of the burst, so if $\lambda$ is much longer than $T$, the gravitational wave signal cannot be considered as ``burst''. To make sure the posterior distribution is normalizable, we assume the hyperprior of $\lambda$ is a proper uniform distribution from $0$ to $T$, 
\begin{equation}
 q_{\lambda}(\lambda) = \frac{1}{T}
\end{equation}

We finally set the hyperprior for the sky location of the source. We expect the sources of gravitational wave bursts to be uniformly distributed across 
the sky, so the prior of the gravitational wave propagation direction $\hat{k}$ is
\begin{equation}
q_{\hat{k}}(\hat{k}) = \frac{1}{4\pi}
\end{equation}
\end{subequations}

\subsection{Inferring $\boldsymbol{\tau}$ and Hyperparameters} \label{sec:inference}
Having chosen the prior probability distribution of $\bm{\tau}$, we need to write down its probability posterior density to make Bayesian nonparametric inference, as described in Sec.~\ref{sec:framework}. Denote the observed timing residuals of the $j$ pulsar as $\mathrm{d}_j$ and the contribution by gravitational wave burst is $\tau_j$, the likelihood function of the $j$th pulsar is
\begin{subequations}
\begin{align}
\Lambda(\mathrm{d}_j|\tau_j) &= N(\mathrm{d}_j-\tau_j|\mathrm{C}_j) \nonumber \\
&=\frac{\exp\left[-\frac{1}{2}(\mathrm{d}_j-\tau_j)^{T}\mathrm{C}^{-1}_j(\mathrm{d}_j-\tau_j)\right]}{\sqrt{(2\pi)^{\mathrm{dim}\,\mathrm{d}_j}\det||\mathrm{C}_j||}}
\end{align}
where $\mathrm{C}_j$ is the noise covariance of the $j$th pulsar. We also assume that the timing noises are uncorrelated among different pulsars. Correspondingly, for a pulsar timing array composed of $N_p$ pulsars, the likelihood function of the timing residuals $\bm{\mathrm{d}}$ of the pulsar timing array is
\begin{align} \label{eq:likelihood}
\Lambda(\bm{\mathrm{d}}|\bm{\tau}) &= \prod^{N_p}_{j=1} \Lambda(\mathrm{d}_j|\tau_j) \nonumber \\
&= N(\bm{\mathrm{d}} - \bm{\tau}|\bm{\mathrm{C}})
\end{align}
\end{subequations}

Following the discussion in Sec.~\ref{sec:inference0} with Eq.~\eqref{eq:likelihood}, Eq.~\eqref{eq:priortau} and Eq.~\eqref{eq:hyperprior}, we can determine the joint posterior probability density of $\bm{\tau}$ and hyperparameters,
\begin{subequations}\label{eq:inference}
\begin{align} \label{eq:posterior}
p(\bm{\tau},\hat{k},\sigma_{+},\sigma_{\times},\lambda|\bm{\mathrm{d}}) &= \frac{1}{Z(\bm{\mathrm{d}})} \Lambda(\bm{\mathrm{d}}|\bm{\tau})q(\bm{\tau}|\hat{k},\sigma_{+},\sigma_{\times},\lambda)q_{\hat{k}}(\hat{k})q_{+}(\sigma_{+})q_{\times}(\sigma_{\times})q_{\lambda}(\lambda) \nonumber \\
&= \sqrt{\frac{\det||\bm{\mathrm{A}}||}{(2\pi)^{\mathrm{dim}\bm{\mathrm{A}}}}}\exp\left[-\frac{1}{2}(\bm{\tau}-\bm{\tau}_{m})^{T}\bm{\mathrm{A}}(\bm{\tau}-\bm{\tau}_{m})\right] \nonumber \\
&\times \frac{1}{Z(\bm{\mathrm{d}})}\Lambda_{\theta}(\bm{\mathrm{d}}|\hat{k},\sigma_{+},\sigma_{\times},\lambda)q_{\hat{k}}(\hat{k})q_{+}(\sigma_{+})q_{\times}(\sigma_{\times})q_{\lambda}(\lambda)
\end{align}
where $\bm{\mathrm{A}}$ is 
\begin{equation}
\bm{\mathrm{A}} = \bm{\mathrm{K}}^{-1} + \bm{\mathrm{C}}^{-1}
\end{equation}
and $\bm{\tau}_m$ satisfies
\begin{equation}
\bm{\mathrm{A}} \bm{\tau}_m = \bm{\mathrm{C}}^{-1} \bm{\mathrm{d}}
\end{equation}
$\Lambda_{\theta}(\bm{\mathrm{d}}|\hat{k},\sigma_{+},\sigma_{\times},\lambda)$ is 
\begin{align} \label{eq:hyperlike}
\Lambda_{\theta}(\bm{\mathrm{d}}|\hat{k},\sigma_{+},\sigma_{\times},\lambda) &= \int \Lambda(\bm{\mathrm{d}}|\bm{\tau})q(\bm{\tau}|\hat{k},\sigma_{+},\sigma_{\times},\lambda) \, \mathrm{d}\bm{\tau} \nonumber \\
&= \frac{\exp\left[-\frac{1}{2}\bm{\mathrm{d}}^{T}\bm{\mathrm{C}}^{-1}\bm{\mathrm{d}}\right]}{\sqrt{(2\pi)^{\mathrm{dim}\,\bm{\mathrm{d}}}\det||\bm{\mathrm{C}}||}} \times \frac{\exp\left[\frac{1}{2}(\bm{\mathrm{C}}^{-1}\bm{\mathrm{d}})^{T}\bm{\mathrm{A}}^{-1}(\bm{\mathrm{C}}^{-1}\bm{\mathrm{d}})\right]}{\sqrt{\det||\bm{\mathrm{A}}||\det||\bm{\mathrm{K}}||}}
\end{align}
\end{subequations}
Eq.~\eqref{eq:inference} summarizes Bayesian nonparametric inference, which gives estimation on the time series function $\bm{\tau}$ and hyperparameters. 

\subsection{Inferring If A Gravitational Wave Burst is Present} \label{sec:DIC}
Given timing residual observations $\bm{\mathrm{d}}$ from an array of pulsars, we would like to infer if a gravitational wave burst is present. We treat this issue as a problem in Bayesian model comparison \cite{gelman:2004:bda}. Consider the two models
\begin{subequations}\label{eq:model}
\begin{align} 
&M_{1} = \left(\text{a gravitational wave burst is present in the data set}\right) \\
&M_{0} = \left(\text{no gravitational waves bursts are present in the data set}\right)  
\end{align}
\end{subequations}
The purpose of model comparison is to check which model data favors. If data favors $M_1$, then it indicates that a 
gravitational wave is likely to be present in the data set. 

Bayes factor, which is fully consistent and derivable from the principle of Bayesian inference, is usually used as the criterion of Bayesian model comparison \cite{kass:1995:bf}. However, Bayes factor is well defined only when the priors and hyperpriors are all proper distributions \cite{kass:1995:bf, gelman:2004:bda}. We can see this point by investigating the definition of Bayes factor (BF), which is the ratio of marginal likelihood functions of the two exclusive hypotheses in \eqref{eq:model}:
\begin{equation}\label{eq:bf}
 \mathrm{BF} = \frac{\int \Lambda_1(\bm{\mathrm{d}}|\bm{\theta}_1, M_1)q_1(\bm{\theta}_1, M_1) \mathrm{d}\bm{\theta}_1}{\Lambda_0(\bm{\mathrm{d}}|M_0)} 
\end{equation}
where $\Lambda_1$ and $\Lambda_0$ are respectively the likelihood functions in model $M_1$ and $M_0$; $\bm{\theta}_1$ denotes all the parameters and hyperparameters in model $M_1$ and there are no parameters or hyperparameters used in the null hypothesis $M_0$. If the prior distribution $q_1(\bm{\theta}_1, M_1)$ is improper, there will be an arbitrary multiplicative constant in Bayes factor, which makes it ill-defined. We may try to transform the improper prior to proper prior by imposing some upper or lower bounds on parameters and make the Bayes factor well defined. However, this procedure will not solve the problem either. Take the two RMS burst amplitudes $\sigma_{+,\times}$ as an example. We can impose an upper bound $\sigma_{m}$ in Eq.~\eqref{eq:hyperpriors} and change both their hyperpriors to uniform distribution from $0$ to $\sigma_{m}$. However, the two new hyperpriors will have a normalization constant $1/\sigma_{m}$ that will appear in Bayes factor Eq.~\eqref{eq:bf}. As a result, Bayes factor will be strongly sensitive to the uncertain cut-off $\sigma_m$ and the evidence of gravitational waves will depend on the subjective choices of $\sigma_m$. 

To avoid the problem of Bayes factor, we decide to use an alternative criterion, {\it Deviance Information Criterion} (DIC) to quantitatively judge how well the data favors a model \cite{spiegelhalter:2002:bmo}. DIC is the sum of two terms --- one term represents ``goodness of fitting'', which measures how well the model fits the data and is the negative log likelihood of the model; the other term represents ``the penalty of complexity'', which measures the degree of overfitting or the effective number of parameters of the model \cite{spiegelhalter:2002:bmo}. The data favors the model with smaller DIC \cite{spiegelhalter:2002:bmo}. In the following subsections, we discuss these two terms for our example problem. 

\subsubsection{Expected Deviance as a Measure of ``Goodness of Fitting''}
In the DIC, the ``goodness of fitting'' is summarized in the so-called ``deviance'', which measures the uncertainty of the model and is defined as $-2$ times the log-likelihood \cite{ kullback:1951:oia, gelman:2004:bda}:
\begin{equation} \label{eq:tdeviance}
 D(\bm{\mathrm{d}},\bm{\tau}) = -2 \log \Lambda(\bm{\mathrm{d}}|\bm{\tau})
\end{equation}
This quantity characterizes the model discrepancy \cite{spiegelhalter:2002:bmo} and resembles the classical $\chi^2$ goodness-of-fit measure. Therefore, the average of 
the deviance on posterior probability distribution provides a summary of the error of model $M_1$ and represents the ``goodness of fitting'' \cite{spiegelhalter:2002:bmo}:
\begin{subequations}
\begin{equation} \label{eq:deviance}
D_{\mathrm{avg}}(\bm{\mathrm{d}},M_1) = \int D(\bm{\mathrm{d}},\bm{\tau}) p_\tau(\bm{\tau}|\bm{\mathrm{d}}) \mathrm{d}\bm{\tau}
\end{equation}
where $p(\bm{\tau}|\bm{\mathrm{d}})$ is the posterior probability density 
in Eq.~\eqref{eq:posterior} marginalizing over all hyperparameters:
\begin{equation} \label{eq:posteriortau}
p_\tau(\bm{\tau}|\bm{\mathrm{d}}) = \int \, p(\bm{\tau},\hat{k},\sigma_{+},\sigma_{\times},\lambda|\bm{\mathrm{d}})\, \mathrm{d}^{2}\mathrm{\Omega}_{k} \mathrm{d}\sigma_{+}\mathrm{d}\sigma_{\times}\mathrm{d}\lambda
\end{equation}
\end{subequations}

For model $M_0$, since there are no parameters representing the model, the average of the deviance is 
\begin{subequations}
\begin{equation}
D_{\mathrm{avg}}(\bm{\mathrm{d}},M_0) = -2 \log \Lambda(\bm{\mathrm{d}}|M_0)
\end{equation}
where $\Lambda(\bm{\mathrm{d}}|M_0)$ is the null model likelihood function
\begin{equation} \label{eq:deviance0}
\Lambda(\bm{\mathrm{d}}|M_{0}) = N(\bm{\mathrm{d}}|\bm{\mathrm{C}}) = \frac{\exp\big(-\frac{1}{2}\bm{\mathrm{d}}^{T}\bm{\mathrm{C}}^{-1}\bm{\mathrm{d}}\big)}{\sqrt{(2\pi)^{\mathrm{dim}\,\bm{\mathrm{d}}}\det||\bm{\mathrm{C}}||}}
\end{equation}
\end{subequations}

\subsubsection{Model Complexity}
Now we need to consider the complexity of a model and the more complex model with more adjustable parameters should have larger 
penalty \cite{gelman:2004:bda, claeskens:2008:msa, jefferys:1991:sor}. For model $M_1$, the parameter set that represents the model is $\bm{\tau}$. In Bayesian analysis, we can use the mean value $\bar{\bm{\tau}}$ of $\bm{\tau}$ under its posterior probability density to be the Bayesian estimator of $\bm{\tau}$ \cite{gelman:2004:bda},
\begin{equation} \label{eq:meantau}
\bar{\bm{\tau}} = \int \, \bm{\tau} \, p_\tau(\bm{\tau}|\bm{\mathrm{d}}) \, \mathrm{d}\bm{\tau}
\end{equation}
We can use it to estimate the uncertainty of model $M_1$ by evaluating deviance $D(\bm{\mathrm{d}},\bm{\tau})$ in Eq.~\eqref{eq:tdeviance}, then the excess of the true over over the estimated uncertainty will be denoted by
\begin{equation}
 \Delta_D(\bm{\mathrm{d}},\bm{\tau},\bar{\bm{\tau}}) = D(\bm{\mathrm{d}},\bm{\tau}) - D(\bm{\mathrm{d}},\bar{\bm{\tau}})
\end{equation}
which can be thought as the reduction in uncertainty due to Bayesian estimation, or alternatively the degree of overfitting due to $\bar{\bm{\tau}}$ adapting to the data set $\bm{\mathrm{d}}$ \cite{spiegelhalter:2002:bmo}. Correspondingly, the mean value of $\Delta_D$ under the posterior probability density summarizes the model complexity or the degree of overfitting due to the model $M_1$ adapting to the data set \cite{spiegelhalter:2002:bmo, meng:1992:plr} 
\begin{align}
 p_D(\bm{\mathrm{d}},M_1) &= \int \, \Delta_D(\bm{\mathrm{d}},\bm{\tau},\bar{\bm{\tau}}) \, p_\tau(\bm{\tau}|\bm{\mathrm{d}}) \, \mathrm{d}\bm{\tau} \nonumber \\
&= D_{\mathrm{avg}}(\bm{\mathrm{d}},M_1) - D(\bm{\mathrm{d}},\bar{\bm{\tau}})
\end{align}
We can see that when the posterior probability density of $\bm{\tau}$ is approximately a Gaussian distribution, $p_D$ is the approximately the trace of the product of Fisher information matrix and the posterior covariance matrix, which is the effective number of parameters \cite{spiegelhalter:2002:bmo}. 
It is reasonable that the model complexity $p_D$ depends on the observed data set because matching the data with the model would induce the statistical correlation among parameters that is likely to reduce the effective dimensionality of the model, and the degree of the reduction may depend on the specific data set \cite{spiegelhalter:2002:bmo}. Correspondingly the degree of overfitting of the model would depend on the data set.

For model $M_0$, since no parameters are needed, so the degree of overfitting for this model, $p_{D}(\bm{\mathrm{d}},M_0)$, is zero.

\subsubsection{Deviance Information Criterion}
We have defined the measure of ``goodness of fitting'' as the mean value of deviance $D_\mathrm{avg}$, which summarizes the uncertainty or the lack of fit of a model. A model with smaller value of $D_\mathrm{avg}$ indicates it fits the data better \cite{kullback:1951:oia}. However, more complex models with more adjustable parameters will usually fit the data better, which opens the possibility of overfitting \cite{gelman:2004:bda}. So we define a measure of degree of overfitting $p_D$ to trade off against $D_\mathrm{avg}$. Based on the discussion above, $p_D$ represents the reduction of lack of the model fit, or the expected improvement in the fit by Bayesian estimation of the parameters in the model \cite{spiegelhalter:2002:bmo}. Correspondingly, the sum of $D_\mathrm{avg}$ and $p_D$ would be the measure of how the data favors the model \cite{spiegelhalter:2002:bmo},
\begin{equation} \label{eq:DIC}
\mathrm{DIC}(\bm{\mathrm{d}},M) =  D_{\mathrm{avg}}(\bm{\mathrm{d}},M) + p_{D}(\bm{\mathrm{d}},M)
\end{equation}
This quantity is referred to as Deviance Information Criterion (DIC). The data would favor the model with smaller DIC, since such a model has smaller discrepancy of the data and is less complex.

The difference between the DICs of two model in Eq.~\eqref{eq:model} ,
\begin{equation}
\Delta \mathrm{DIC} = \mathrm{DIC}(\bm{\mathrm{d}},M_{1}) - \mathrm{DIC}(\bm{\mathrm{d}},M_{0})
\end{equation}
characterizes how much the timing residual observations favors $M_1$ over $M_0$. It is similar to 
negative twice the natural logarithm of Bayes factor \cite{kass:1995:bf}. Correspondingly, it implies that $\exp(-\Delta\mathrm{DIC}/2)$ has the same scale as Bayes factor \cite{spiegelhalter:2002:bmo}. If $\Delta \mathrm{DIC} \lesssim -10$, it implies that the equivalent Bayes factor between the two models is $\sim 150$. So it is safe to conclude that the data 
strongly favors $M_{1}$ and there is strong evidence that a gravitational wave burst is present in the data set \cite{spiegelhalter:2002:bmo}.

\subsection{Comparison with F\&L} \label{sec:fl}
At this point, it is worth comparing our Bayesian nonparametric method described above and the popular method F\&L. 

\subsubsection{Incorporation of Gravitational Waveform Characteristics} 
In our Bayesian nonparametric method, we incorporate the expected characteristics of gravitational waveforms into a Gaussian process prior to set strong constraints on the feasible signal patterns (see Section \ref{sec:priorgw}). Especially, due to the smoothness of the gravitational waveforms as expected, we are able to narrow down our searches for {\it smooth} signals only by setting the prior Eq.~\eqref{eq:priortau}. As a bonus, we can estimate the burst durations even though we do not have physical models for the sources. This is exactly the strength of Bayesian nonparametric analysis \cite{rasmussen:2006:gpf}.

In comparison, F\&L does not incorporate any gravitational waveform characteristics into their analysis, because they ignore the temporal correlation and the smoothness of the gravitational waveforms by choosing the kernels in the Gaussian process prior as diagonal matrices \cite{finn:2010:dla}. Such kernels do not contain the information of burst durations and the corresponding Gaussian process priors do not set any physical constraints or expected characteristics on $\tau_{(+)}$ and $\tau_{(\times)}$. Correspondingly, their analysis would offer redundant degrees of freedom and provide less informed inference. Therefore, even though F\&L also tries to directly infer the burst signals but not fit the data with a specific physical model, it may not be considered as Bayesian nonparametric analysis because it misses the most important advantage of such analysis. 

\subsubsection{Marginalization over Hyperparameters}
In our Bayesian nonparametric analysis, we choose prior distributions for both parameters and hyperparameters, and we compute the posterior distributions and DICs by marginalizing over parameters and hyperparameters to perform a full Bayesian inference. In this way, we provide a robust inference and minimize the risk of false alarms and overfitting \cite{gelman:2004:bda}. 

In comparison, F\&L does not assign any prior distributions on hyperparameters and it maximizes the likelihood function over them to infer parameters and compute Bayes factor \cite{finn:2010:dla}. This optimization procedure is only a good approximation when the hyperparameters are precisely determined \cite{rasmussen:2006:gpf}. Because in this case, the likelihood function on hyperparameters, i.e., Eq.~\eqref{eq:hyperlike}, is sharply peaked and averaging it over hyperparameters would be approximately equal to maximizing it over hyperparameters. However, gravitational waves searched by pulsar timing arrays are generally very weak, and the hyperparameters are not well measured. Maximizing likelihood function over hyperparameters will significantly overestimate Bayes factor and underestimate the parameter uncertainties. This point is indicated in Eq.~\eqref{eq:bf}, the definition of Bayes factor. The numerator in Eq.~\eqref{eq:bf} will be maximum likelihood if we do not assign any prior distributions and maximize the likelihood function over the hyperparameters as F\&L. The maximum likelihood is always larger than the averaged likelihood over parameters as in Eq.~\eqref{eq:bf}, and the difference will be much greater if the parameters are not precisely measured. This is the reason why F\&L can obtain an incredily huge Bayes factor $\sim\exp(66)$ even if the signal is very weak with an amplitude signal-to-noise ratio less than $1$ (see Table~2 in \cite{finn:2010:dla}). Therefore, the optimization procedure used in F\&L may significantly raise the risk of false alarm and underestimate the parameter uncertainties. We will demonstrate this point in Section \ref{sec:demo}.



\section{Examples}\label{sec:demo}
\subsection{Overview}
To illustrate the effectiveness of the analysis techniques just described, we apply them to simulated observations of a gravitational wave burst generated by a periapsis passage of a long period supermassive black hole binary in highly eccentric orbit. We consider 3 cases:
\begin{itemize}
 \item[(1)] a strong signal, in which we can not only detect the signal, but also localize the source in the sky.
 \item[(2)] a weak signal, in which we are able to detect the signal, but not able to accurately localize its source.
 \item[(3)] no signal at all. 
\end{itemize}
We will also apply F\&L on these three examples and compare the results with our analysis. 

For these examples, we use 4 pulsars in the current International Pulsar Timing Array (IPTA) \cite{hobbs:2010:tip, demorest:2012:lot, manchester:2012:tpp} which are most accurately timed as described in Table~\ref{tab:IPTA}. The capability of detecting and characterizing gravitational waves is dominated by these best pulsars, although they are the minority of the full IPTA \cite{burt:2011:opt}. The timing noises of those 4 pulsars are superposition of short-timescale white noise with rms timing residual given in Table~\ref{tab:IPTA} and long-timescale red noise normalized to have the same spectral density as the white noise at frequency 0.2 $\mathrm{yr}^{-1}$. 

\begin{table}[ht]
\centering
\caption{\hspace{0.3cm} 4 IPTA pulsars we use, Their white timing noise rms and the Telescopes from which the timing residuals are measured \cite{hobbs:2010:tip, demorest:2012:lot, manchester:2012:tpp}}
\vspace{0.5cm}
\begin{tabular}{c c c}
\hline \hline
Pulsar \hspace{1.5cm} & RMS Residual (ns) & \hspace{1.5cm} Telescope \\
\hline
J1713$+$0747 \hspace{1.5cm} & 30 &  \hspace{1.5cm} AO \\
J1909$-$3744 \hspace{1.5cm} & 38 & \hspace{1.5cm} GBT \\
J0437$-$4715 \hspace{1.5cm} & 75 & \hspace{1.5cm} Parkes \\
J1857$+$0943 \hspace{1.5cm} & 111 & \hspace{1.5cm} AO \\
\hline
\end{tabular}
\label{tab:IPTA}
\end{table}

The data sets we use for these examples are constructed by 
\begin{itemize}
 \item[(1)] evaluating the pulsar timing response of each pulsar to a passage of a gravitational wave generated by a periapsis passage of a long period supermassive black hole binary in highly eccentric orbit (See Sec.~\ref{sec:signal}).
\item[(2)] adding the pulsar timing noise to the timing residuals obtained by the first step (See Sec.~\ref{sec:noise}).
\item[(3)] removing the best-fit linear trend from the noisy timing residuals obtained by the second step because in reality, the linear trend in the data cannot be distinguished from the systematic effects caused by pulsar spin and spin down \cite{hobbs:2006:tan, edwards:2006:tan}.
\end{itemize}

\subsection{Construction of Simulated Data Sets}
\subsubsection{Periapsis Passage of a Long Period Supermassive Black Hole Binary in Highly Eccentric Orbit}\label{sec:signal}

\begin{figure}[ht]
\caption{\hspace{0.1cm} Prefit and Postfit Pulsar Timing Residuals induced A Gravitational Wave Burst}
\centering
\includegraphics[width=15cm]{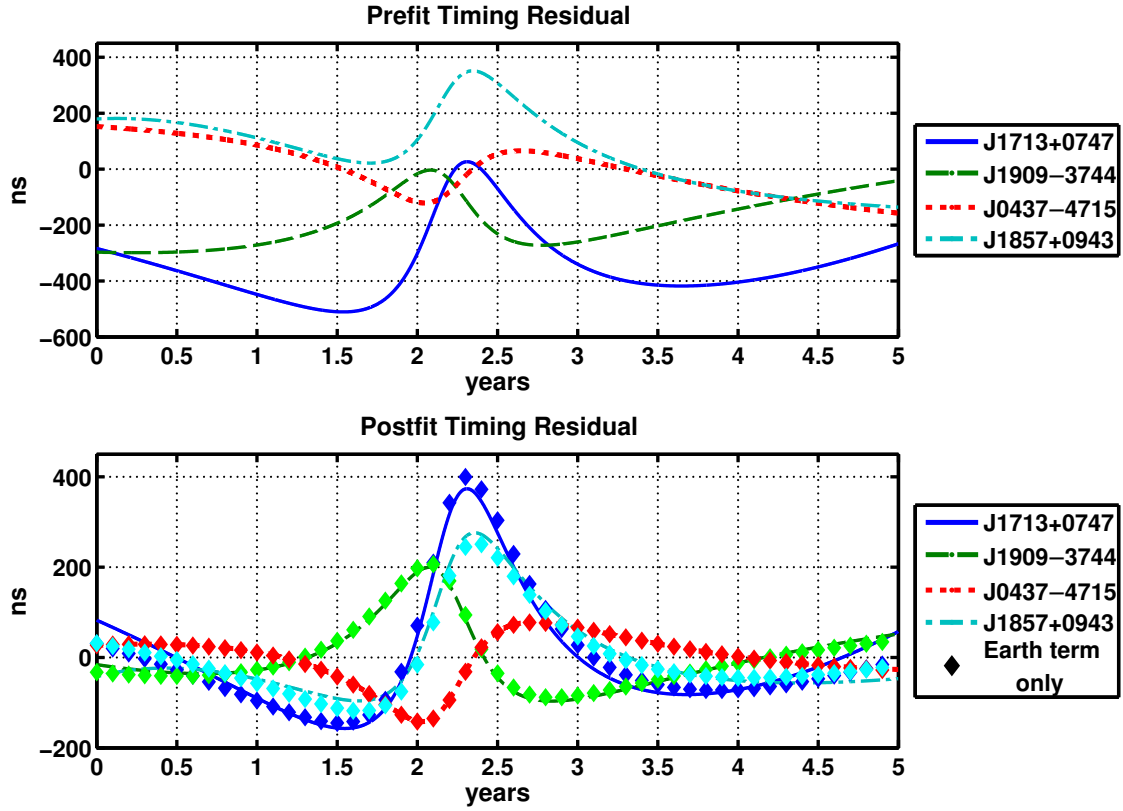}
\begin{flushleft}
{\bf Notes.} In this example, the gravitational wave burst is generated by a periapsis passage of a supermassive black hole binary with total mass of $2\times10^9\,\mathrm{M_\odot}$, symmetric mass ratio of $0.2$, period of $20$ years, eccentricity of $0.8$, orbital inclination angle of $30^{\circ}$, and located at Virgo Cluster. The upper panel shows the timing residuals of the 4 IPTA pulsars induced by this burst, and the lower panel shows the same timing residuals but after ``fitting-out'' the linear trend. The ``diamond'' plot in the lower panel shows the timing residuals only considering Earth term contributions, and we can see that they are approximately the same as the post-fit timing residuals. 
\end{flushleft}

\label{fig:signals}
\end{figure}

Our hypothetical gravitational wave burst source is a periapsis passage of a supermassive black hole binary with total mass of $2\times10^9\,\mathrm{M_\odot}$, symmetric mass ratio of $0.2$, period of $20$ years, eccentricity of $0.8$, and orbital inclination angle of $30^{\circ}$. The duration of this burst is $0.42\,\mathrm{yr}$, estimated as twice the ratio of the impact parameter to the velocity at periapsis \cite{finn:2010:dla}. It is in the direction of Virgo cluster ($\mathrm{RA}\,12\mathrm{h}27\mathrm{m},\, \mathrm{dec}\,12^{\circ}43'$) and we change the luminosity distance to obtain the strong and weak signals. We randomly sample 50 observation times that are uniformly distributed across the 5 year observations for each pulsar. Fig.~\ref{fig:signals} shows the gravitational wave induced timing residuals of 4 pulsars when the source is at a distance of $16.5\,\mathrm{Mpc}$ (at Virgo cluster). 

\subsubsection{Pulsar Timing Noise}\label{sec:noise}
The millisecond pulsars used in current International pulsar timing array typically show white noise on short timescales, and 
few of them turn to red noise on timescales $\gtrsim 5$ years \cite{demorest:2012:lot, manchester:2012:tpp}. For demonstrations here, we model the timing noise as the superposition of white noise and red noise, with the red noise contribution normalized to have the same amplitude at the white 
noise contribution at the frequency $f_{r}=0.2\,\mathrm{yr^{-1}}$. The power spectral density is taken to be \cite{coles:2011:pta}
\begin{subequations}
\begin{equation}\label{eq:noise}
S_{n}(f) = \sigma^{2}_{n}+\sigma^{2}_{n}\left[\frac{1+\left(\frac{f}{f_0}\right)^2}{1+\left(\frac{f_r}{f_0}\right)^2}\right]^{-5/2}
\end{equation}
where
\begin{align}
\sigma_{n} &= \left(\text{white noise rms}\right) \\
f_r & = \left(\text{red-white noise cross-over frequency, $0.2\,\mathrm{yr^{-1}}$}\right)
\end{align}
\end{subequations}
and $f_0$ softens the noise spectrum at ultra-low frequency. As long as $f_0$ is much less than the pulsar timing array frequency 
band, its value does not matter. In the simulation we set $f_0$ equal to $0.01\mathrm{nHz}$. We choose the power index of the red noise 
spectrum as $-5$ because the few millisecond pulsars showing red noises have noise spectrum with power index $-5$ \cite{shannon:2010:atr}. 

The covariance matrix of the noise will be the Fourier transform of the noise specturm density Eq.~\eqref{eq:noise} to time 
domain, i.e., 
\begin{equation} \label{eq:cov}
\mathrm{C}(t_i,\, t_j) = \sigma^{2}_{n}\left(\delta_{ij} + \sqrt{\frac{2}{9\pi}} \left[1+\left(\frac{f_r}{f_0}\right)^2\right]^{5/2}f^3_0(t_i-t_j)^{2}K_{2}(f_0|t_i-t_j|)\right)
\end{equation}
where $t_{i,j}$ are the ``observation times'' and $K_2$ is the modified Bessel function of the second kind with index $2$. The pulsar 
timing noises for each pulsar are sampled from multivariate normal distribution with zero mean and covariance matrix Eq.~\eqref{eq:cov}.

\subsection{Analysis of Simulated Data Sets}
Our Bayesian nonparametric analysis is designed to investigate if a gravitational wave burst is present in the simulated dataset, and also infer the source sky location, the burst duration and other hyperparameters. We use Markov Chain Monte Carlo method \cite{robert:2004:mcs} to compute the posterior probability densities and Deviance Information Criterion described in Sec.~\ref{sec:inference} and Sec.~\ref{sec:DIC}. We have found that for weak signals, it may be able to clearly detect the gravitational wave but not able to accurately localize the sources or infer $\bm{\tau}$; but for strong signals, it may be possible to both detect the gravitational wave and also precisely localize the sources and infer $\bm{\tau}$. We illustrate this point in the following three subsections, first investigating the strong signal example, then weak signal counterpart, and finally, we analyze the dataset consisting of timing noises alone for comparative study. In each of these three cases, we also apply F\&L to analyze the simulated dataset, but instead of maximizing likelihood, we marginalize over hyperparameters to perform a robust inference. The results are summarized in Table~\ref{tab:results} and those obtained by F\&L are listed in the parentheses in the second and third column. The signal-to-noise ratio (SNR) is defined as that in F\&L \cite{finn:2010:dla}, i.e., 
\begin{equation}
\mathrm{SNR}^2 = \bm{\tau}_0 \bm{\mathrm{C}}^{-1} \bm{\tau_0}
\end{equation}
where $\bm{\tau}_0$ denotes the ``actual'' signal and $\bm{\mathrm{C}}^{-1}$ is the noise covariance matrix.

\begin{table}[ht]
\centering
\caption{\hspace{0.1cm} Results for Bayesian Nonparametric Analysis on 3 Simulated Data Sets and Comparison with F\&L (parentheses in second and third column)}
\vspace{0.5cm} 
\begin{tabular}{c c c c c c}
\hline \hline
SNR \hspace{1cm} & $\Delta\mathrm{DIC}$ \hspace{1cm} & $\Delta\Omega_{k}\,\mathrm{(deg^2)}$ \hspace{1cm} & $\epsilon_{\sigma_{+}}$  \hspace{1cm} & $\epsilon_{\sigma_{\times}}$ \hspace{1cm} & $\epsilon_{\lambda}$\\
\hline
30  \hspace{1cm} & -956 (-245) \hspace{1cm} &  3600 (8230) \hspace{1cm} & $23.6\%$ \hspace{1cm} & $32.4\%$ \hspace{1cm} & $11.4\%$\\
5  \hspace{1cm} & -14 (-4) \hspace{1cm} &  26702 (36205) \hspace{1cm} & $88.1\%$  \hspace{1cm} &  $112.4\%$ \hspace{1cm} & $50.0\%$\\
0  \hspace{1cm} & 5 (4) \hspace{1cm} & 40840 (40840) \hspace{1cm} & $174.0\%$  \hspace{1cm} &  $232.0\%$ \hspace{1cm} & $58.9\%$\\
\hline
\end{tabular}
\begin{flushleft}
{\bf Notes.} In all cases, the signal corresponds to gravitational wave burst generated from a periapsis passage of a supermassive black hole binary with total mass of $2\times10^9\,\mathrm{M_\odot}$, symmetric mass ratio of $0.2$, period of $20$ years, eccentricity of $0.8$, and orbital inclination angle of $30^{\circ}$, propagating from the direction of Virgo cluster. The source is placed at $16.5\mathrm{Mpc}$ for simulating strong signal and at $100\mathrm{Mpc}$ for simulating weak signal. $\Delta\Omega_{k}$ denotes the measured uncertainty of the sky location of the source, which takes $99\%$ of the total probability; $\epsilon_{\sigma_{+}}$, $\epsilon_{\sigma_{\times}}$ and $\epsilon_{\lambda}$ respectively denote the fractional error of $\sigma_{+}$, $\sigma_{\times}$ and $\lambda$ (fractional error is defined as the measured uncertainty that takes $68.5\%$ of the total probability over the mean value of the measured parameter). The results of F\&L are presented in the parentheses in the second and third column, which are calculated by marginalizing but not optimizing over hyperparameters (see main text for details).
\end{flushleft}
\label{tab:results}
\end{table}

\subsubsection{Strong Signal} \label{sec:strong}
We simulate the strong signal by placing the source described in Sec.~\ref{sec:signal} at a distance of $16.5\mathrm{Mpc}$ (at Virgo Cluster). The peak gravitational wave induced timing residuals of the ``$+$'' and ``$\times$'' polarization components (amplitude of $\tau_{(A)}$ in Eq.~\eqref{eq:tauP}) are respectively $133\,\mathrm{ns}$ and $131\,\mathrm{ns}$. The duration of this burst, which may be estimated as twice the ratio of the impact parameter to the velocity at the periapsis of the elliptical orbit \cite{finn:2010:dla}, is about $0.42\,\mathrm{yr}$. We apply our Bayesian nonparametric analysis described in Sec.~\ref{sec:bayes} to this ``strong signal'' data set (see Fig.~\ref{fig:ds}), and the first row of Table.~\ref{tab:results} and Fig.~\ref{fig:ssky} and \ref{fig:shyper} summarize the results:

\begin{figure}
\includegraphics[width=15cm]{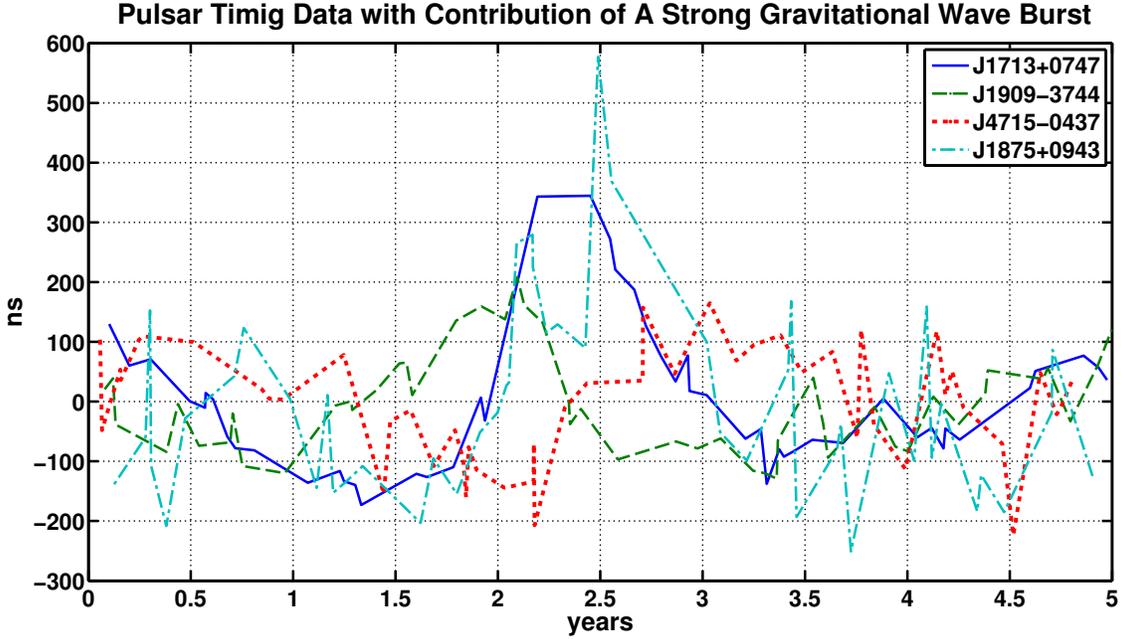}
\caption{''Strong Signal'' Data Set}
\label{fig:ds}
\end{figure}

\begin{itemize}
\item[(1)] From the first row of Table.~\ref{tab:results}, we see that the difference between the DICs of the positive hypothesis and null hypothesis, described in Sec.~\ref{sec:DIC}, is $-956$, corresponding to decisive evidence for the presence of a gravitational wave burst in the data set. Applying F\&L leads to a DIC difference of $-245$ and so offers less decisive evidence of the burst. For demonstration purpose, We also maximize likelihood as in \cite{finn:2010:dla}. In this way, the Bayes factor of our analysis is $\exp(8600)$ and that of F\&L is $\exp(5700)$. However, as we discussed in Section \ref{sec:fl}, the optimization procedure can significantly overestimate the Bayes factor. 
\item[(2)] Having concluded that a signal is present, we use the Bayesian nonparametric inference desccribed in Sec.~\ref{sec:inference} to infer the sky location of the gravitational wave source. Fig.~\ref{fig:ssky} shows the posterior probability density of the sky location $\hat{k}$ marginalizing over all possible $\bm{\tau}$ and all other hyperparameters. The area corresponding to $99\%$ of the total probability is $3600\,\mathrm{deg^2}$, and the actual sky location of the source is within this area. F\&L offers a less precise measurement with a $99\%$ of the total probability $8230\,\mathrm{deg^2}$. We will obtain $\ll 1\,\mathrm{deg^2}$ uncertainty for both our method and F\&L if we maximize the likelihood, but this procedure may significantly underestimate the parameter uncertainties.

\begin{figure}[ht]
\centering
\includegraphics[width=15cm]{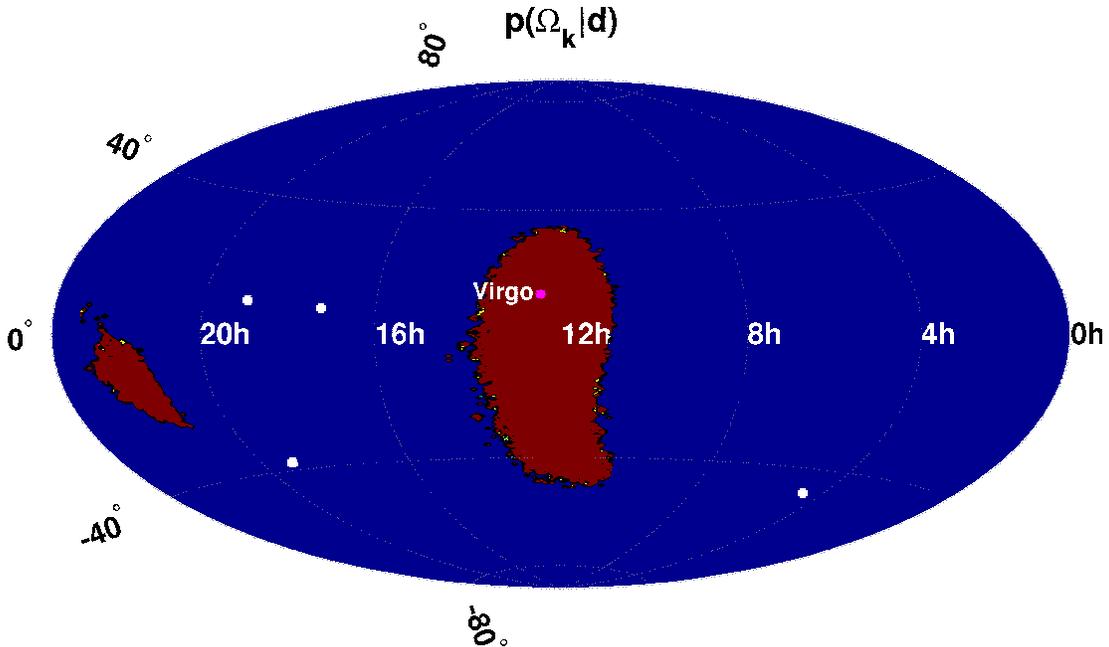}
\caption{Posterior probability density that the source is found at location $\Omega_{k}$ for analysis on ``strong signal'' data set. The inferred sky location has $99\%$ of probability staying within the red region, and the actual sky location is labelled by ``Virgo''. The white squares show the locations of the 4 IPTA pulsars used as our pulsar timing array. See main text for details.}
\label{fig:ssky}
\end{figure}

\item[(3)] We can also infer other hyperparameters such as rms gravitational wave amplitude $\sigma_{+,\,\times}$ and characteristic length-scale $\lambda$. Fig.~\ref{fig:shyper} shows the posterior probability density of these hyperparameters, marginalized over all possible $\bm{\tau}$ and sky locations. 
%

\begin{figure}[ht]
\centering
\includegraphics[width=15cm]{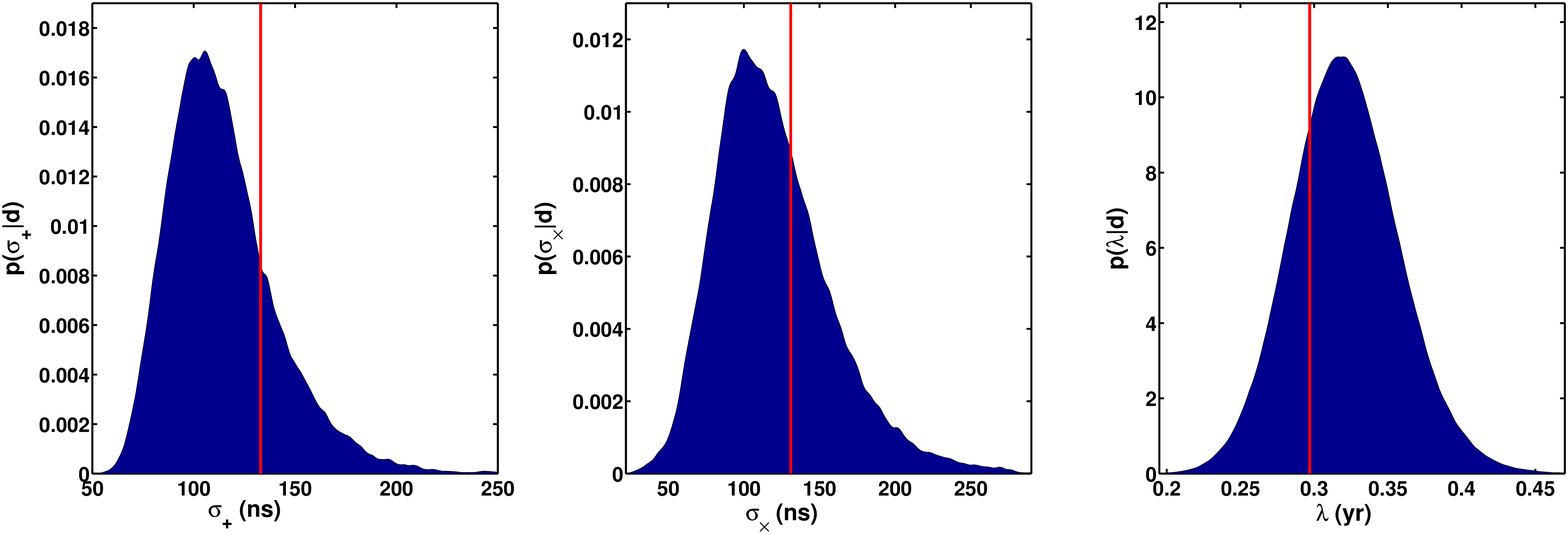}
\caption{Posterior probability densities of 3 hyperparameters --- $\sigma_{+}$, $\sigma_{\times}$ and $\lambda$, for analysis on the ``strong signal'' data described in Sec.~\ref{sec:strong}. In the left panel, it shows the fractional error of $\sigma_{+}$ is about $23.7\%$, and the red line represent the peak gravitational wave induced timing residual of the ``$+$'' polarization component of the simulated source. In the middle panel, it shows the fractional error of $\sigma_{\times}$ is about $32.4\%$, and the red line represents the peak gravitational wave induced timing residual of the ``$\times$'' polarization component of the simulated source. In the right panel, it shows the fractional error of $\lambda$ is about $11.4\%$, and the red line represents the duration of the simulated burst divided by $\sqrt{2}$.}
\label{fig:shyper}
\end{figure}
\end{itemize}
Therefore, for this strong signal, we can find decisive evidence of its presence in the simulated data, and also we can localize its source and estimate its burst duration well. 



\subsubsection{Weak Signal} \label{sec:weak}
To simulate a weak signal, we place the gravitational wave source at a distance of $100\,\mathrm{Mpc}$ but still keep it in the direction of Virgo Cluster. The peak gravitational wave induced timing residuals of the two polarization components are respectively $21.8\,\mathrm{ns}$ and $21.6\,\mathrm{ns}$. We apply the Bayesian nonparametric analysis to this ``weak signal'' data set (see Fig.~
\ref{fig:dw}), and the second row of Table.~\ref{tab:results} and Fig.~\ref{fig:wsky} and \ref{fig:whyper} summarize the results:

\begin{figure}
\includegraphics[width=15cm]{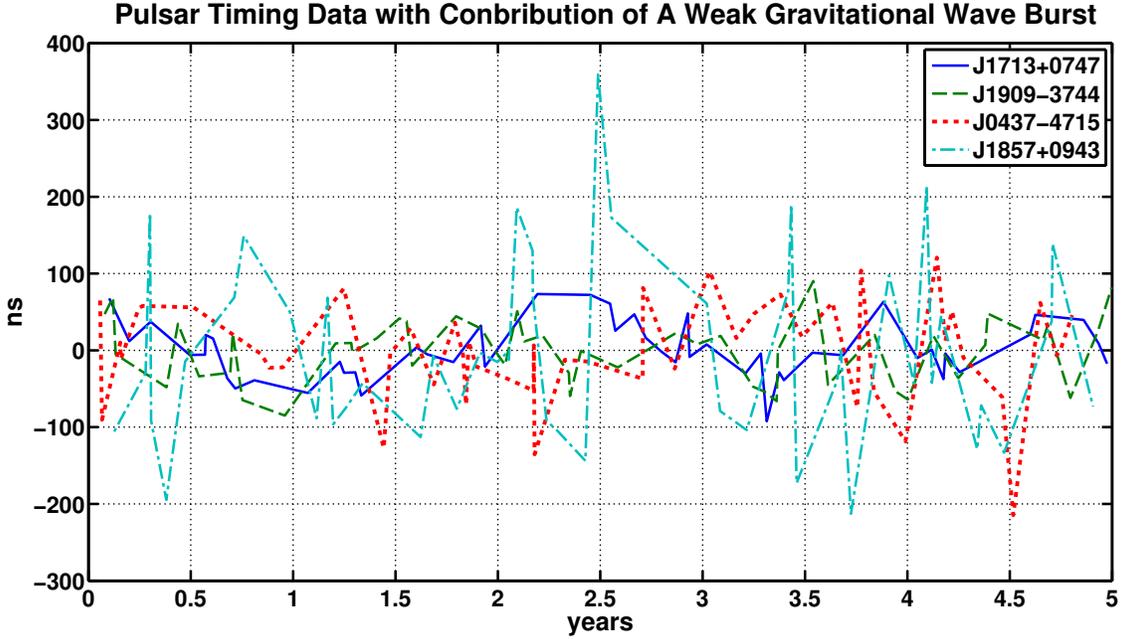}
\caption{''Weak Signal'' Data Set}
\label{fig:dw}
\end{figure}

\begin{itemize}
 \item[(1)] From the second row of Table.~\ref{tab:results}, we see that the difference between the DICs of the two exclusive hypothesis is $-14$, which implies a strong evidence of the presence of a gravitational wave burst in the data set. However, F\&L only offers a DIC difference of $-4$, which implies no strong evidence of the burst. We also maximize the likelihood as in \cite{finn:2010:dla} for demonstration. In this way, the Bayes factor obtained by our method is $\exp(5600)$ and that obtained by F\&L is $\exp(3200)$. Such incredibly huge Bayes factors for weak signals result from the incorrect approximation of maximum likelihood procedure.
\item[(2)] Having obtained a strong evidence of the presence of signal, we begin to infer the sky location of the source. Fig.~\ref{fig:wsky} shows the posterior probability density of the sky location. The area that contains $99\%$ of the total probability is about $26702\,\mathrm{deg}^2$. The error computed by F\&L is $36205\,\mathrm{deg}^2$. If we maximize the likelihood, both our method and F\&L offer the sky location error $\ll 1\,\mathrm{deg}^2$ even for this weak signal, so the optimization procedure significantly underestimate the error of sky location.

\begin{figure}[ht]
\centering
\includegraphics[width=15cm]{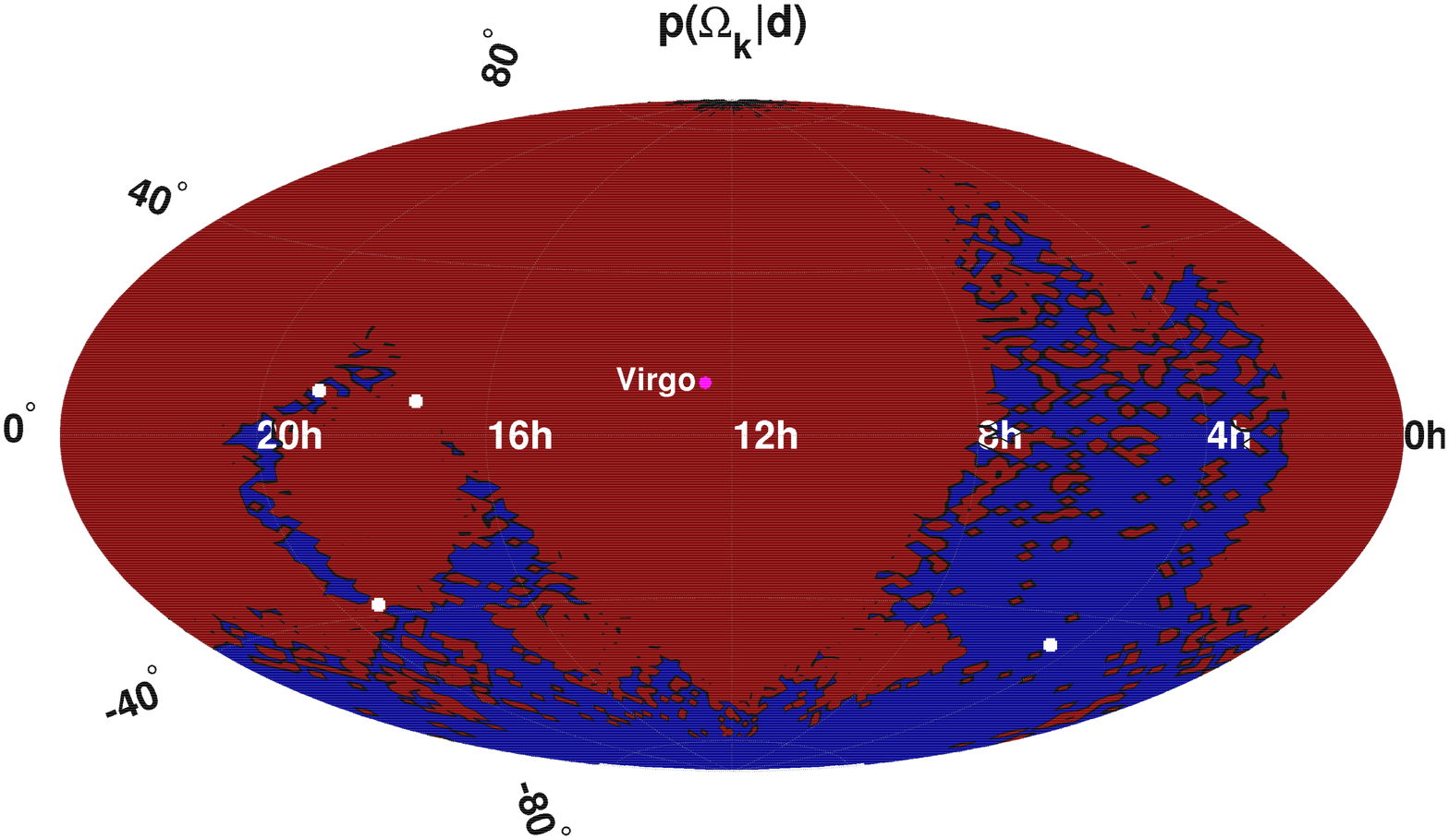}
\caption{Posterior probability density that the source is found at location $\Omega_{k}$ for analysis on ``weak signal'' data set. The inferred sky location has $99\%$ of probability staying within the red region, and the actual sky location is labelled by ``Virgo''. The white squares show the locations of the 4 IPTA pulsars used as our pulsar timing array. See main text for details.}
\label{fig:wsky}
\end{figure}

\item[(3)] We also make inference on other hyperparameters. Fig.~\ref{fig:whyper} shows the posterior probability densities of $\sigma_{+,\,\times}$ and $\lambda$. 
%

\begin{figure}[ht]
\centering
\includegraphics[width=15cm]{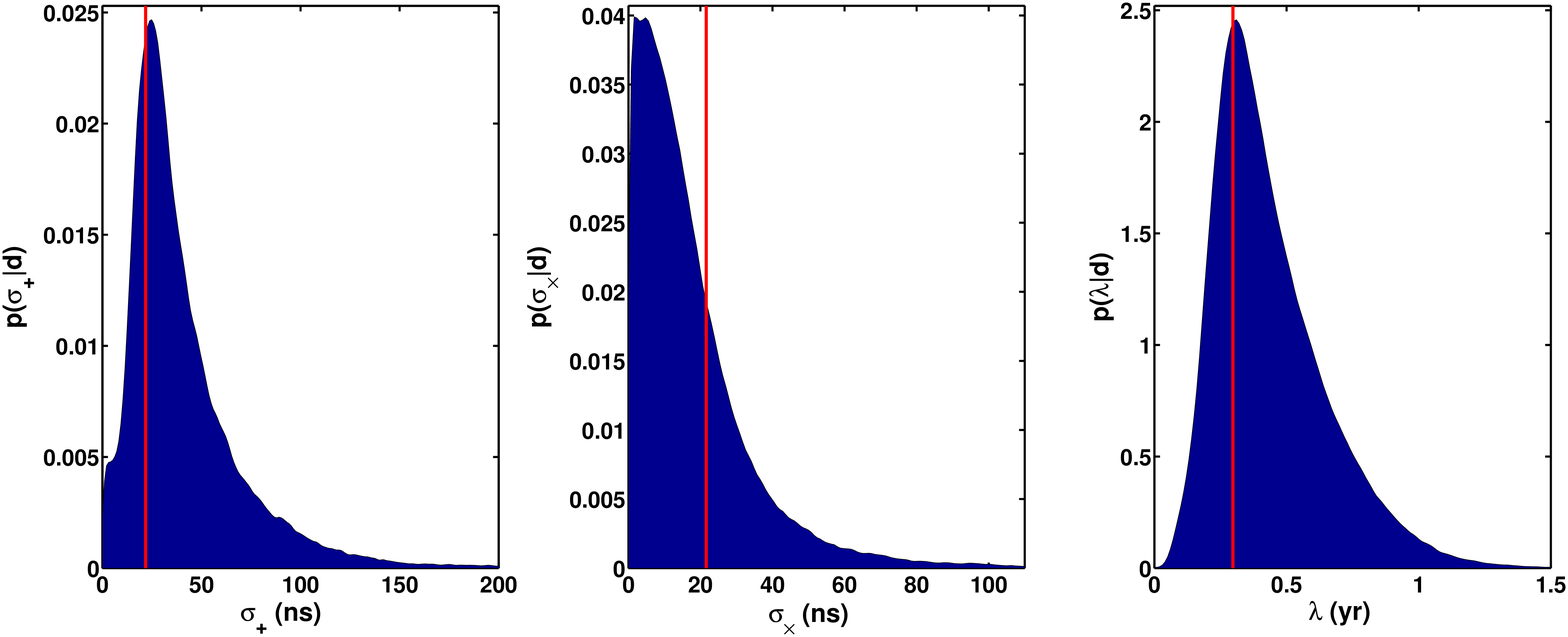}
\caption{Posterior probability densities of 3 hyperparameters --- $\sigma_{+}$, $\sigma_{\times}$ and $\lambda$, for analysis on the ``weak signal'' data described in Sec.~\ref{sec:weak}. In the left panel, it shows the fractional error of $\sigma_{+}$ is about $88.1\%$, and the red line represents the peak gravitational wave induced timing residual of the ``$+$'' polarization component of the simulated source. In the middle panel, it shows the fractional error of $\sigma_{\times}$ is beyond $100\%$ and the red line represents the peak gravitational wave induced timing residual of the ``$\times$'' polarization component of the simulated source. In the right panel, it shows the fractional error of $\lambda$ is about $50\%$, and the red line represents the duration of the simulated burst divided by $\sqrt{2}$.}
\label{fig:whyper}
\end{figure}
\end{itemize}
Therefore, for this weak signal, we can only marginally detect it and make crude inference.



\subsubsection{No Signal} \label{sec:none}
\begin{figure}[ht]
\centering
\includegraphics[width=15cm]{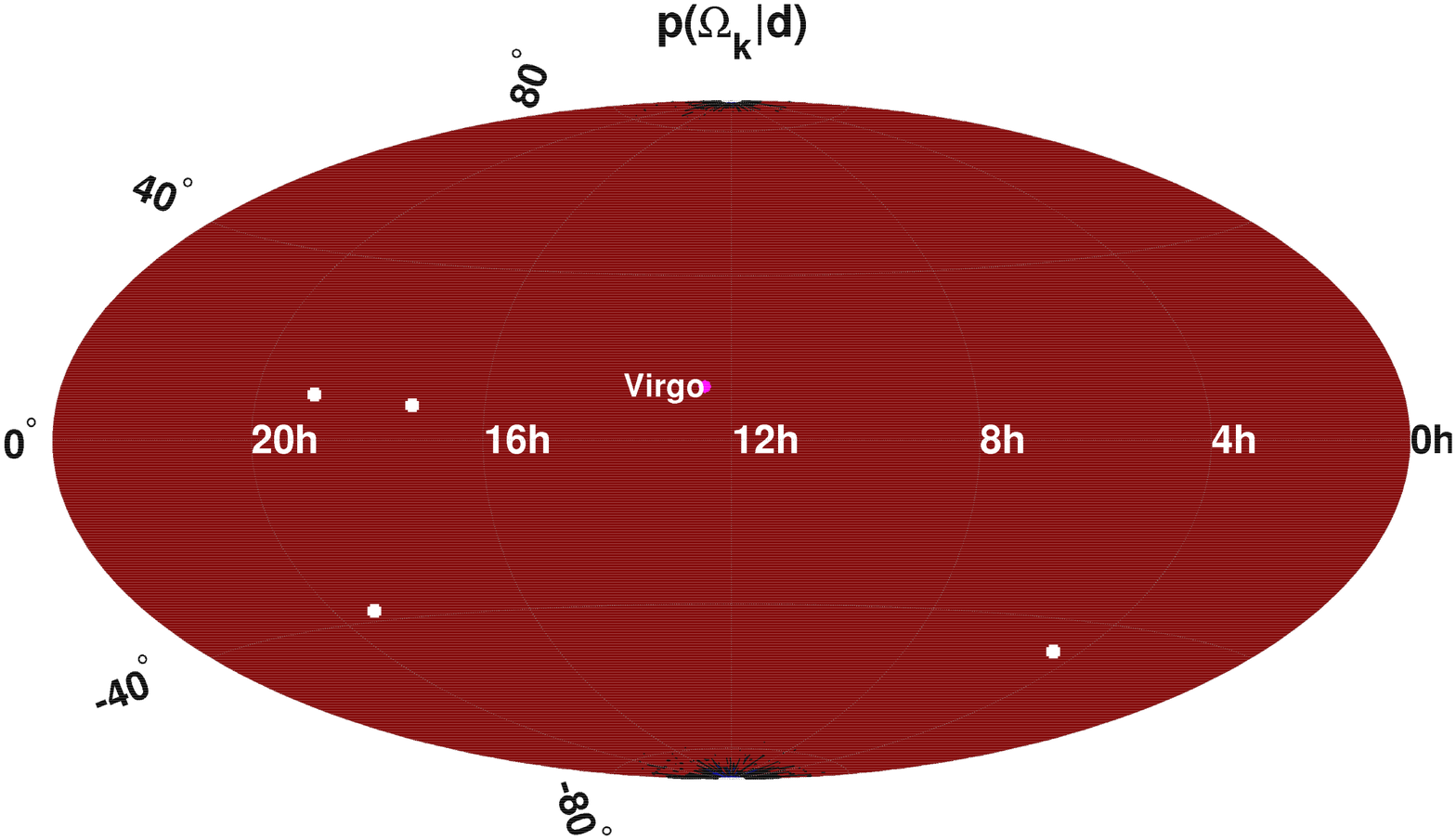}
\caption{Posterior probability density that the source is found at location $\Omega_{k}$ for analysis on ``noise alone'' data set. The inferred sky location has $99\%$ of probability staying within the red region. The white squares show the locations of the 4 IPTA pulsars used as our pulsar timing array. See main text for details.}
\label{fig:nsky}
\end{figure}

\begin{figure}[ht]
\centering
\includegraphics[width=15cm]{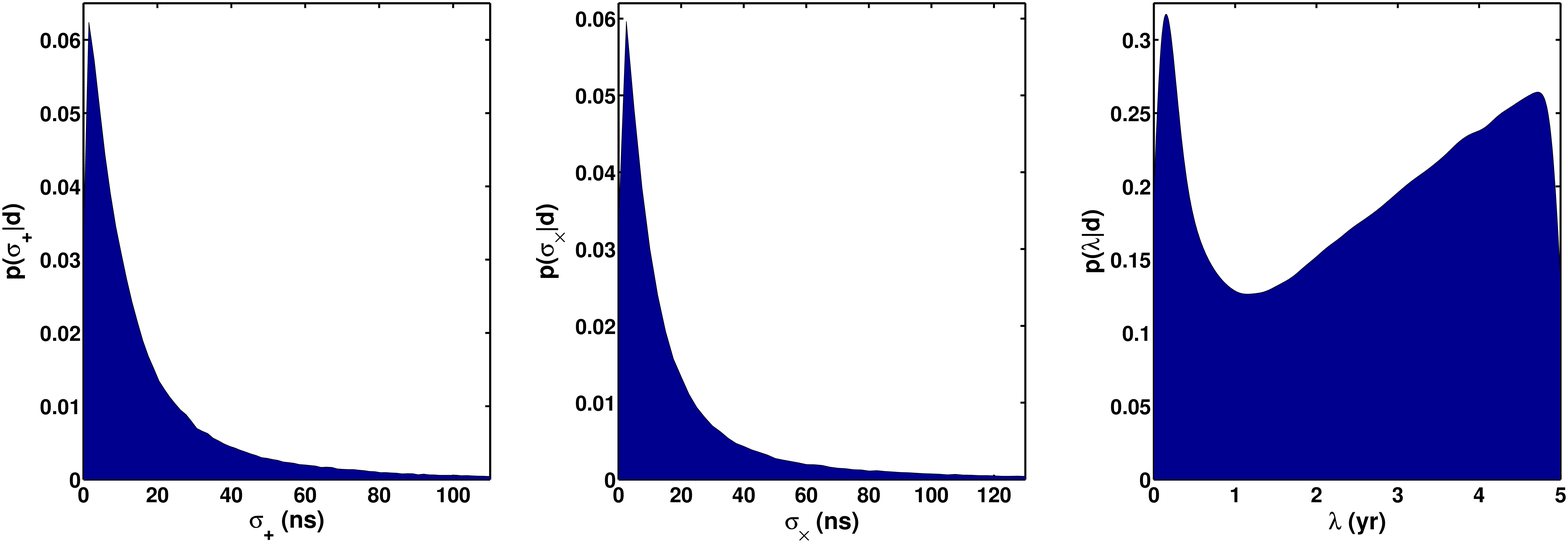}
\caption{Posterior probability densities of 3 hyperparameters --- $\sigma_{+}$, $\sigma_{\times}$ and $\lambda$, for analysis on the ``noise alone'' data described in Sec.~\ref{sec:none}. }
\label{fig:nhyper}
\end{figure}

For comparative study, we also apply our Bayesian nonparametrics analysis to a data set with timing noises alone. The third row of Table.~\ref{tab:results} and Fig.~\ref{fig:nsky} - \ref{fig:nhyper} summarize the results. The difference between the DICs of the two exclusive hypothesis is $5$, which shows that data favors the null hypothesis. The inference of sky location and other hyperparameters has no connection with the source described in Sec.~\ref{sec:signal}. F\&L offers the similar results for the ``no signal'' case. 

%
%

\section{Conclusions}\label{sec:concl}
First detection of gravitational waves will open a new window of our universe complementary with the conventional electromagnetic astronomy. Observing this new window will benefit from advanced analysis methodology and techniques. In this paper, we use a Baysian nonparametric method to analyze the pulsar timing array data set which may contain contribution from a gravitational wave burst. We have investigated how this technique can be used to determine if a gravitational wave burst is present in the data and, if so, infer the sky location of the source and the duration of the burst.

Even though we may not know exactly the analytical formulae of the waveform of the bursts, we still have some expectations of the properties of the bursts: (1) it should be a smooth function of the time; (2) it should have a characteristic duration which should be shorter than the observation duration.  By using the novel Bayesian nonparametrics method, we characterize these two important properties of the gravitational wave bursts into the Gaussian process prior Eq.~\eqref{eq:priortau}. Correspondingly we are able to detect gravitational wave bursts which are too weak to allow their sources to localized, and make accurate inferences on the sky location of the sources and the shape of the signals when the sources are strong. Compared with the analysis in Finn \& Lommen \cite{finn:2010:dla}, we improves the detection sensitivity and provide robust inference. We also infers additional important information such as the duration of the burst. This is because they ignored the temporal correlation of the gravitational waveform, and they cannot incorporate the important characteristics of the bursts described above into their analysis.   

For the purpose of demonstration, we apply our Bayesian nonparametrics analysis to the pulsar timing data of the 4 best millisecond pulsars in current International pulsar timing array (IPTA), as the capability of detection and characterization of gravitational waves will be dominated by these pulsars \cite{burt:2011:opt}. However, our analysis can be straightforwardly applied to analyze the data of all the pulsars in IPTA. In the future, the effective number of pulsars whose timing noises are low enough to detect gravitational waves is expected to signicantly increase with the birth of more sensitive radio telescopes such as Five-hundred-meter Aperture Spherical Telescope \cite{nan:2011:tfa} and Square Kilometer Array (SKA) \cite{dewdney:2009:tsk}. Applying our analysis method to the pulsar timing data accumulated by these future telescopes will significantly improve the detection sensitivity and inference of the sky location of the sources.

While the context of our discussion is gravitational wave burst detection via pulsar timing arrays, following the discussion in Sec.~\ref{sec:bayes} and choosing an appropriate Gaussian process priors with an appropriate kernel, our analysis itself can be directly applied to detecting any kinds of signals whose analytical formulae are unknown, such as detection of binary merger by LIGO \cite{aasi:2013:sfg}, detection of dynamical chaos in exoplanetary systems by Kepler Mission \cite{deck:2012:rdc}, detection of ultra-high energy gamma rays from Gamma Ray Bursts driven by magnetohydrodynamics by Fermi satellite \cite{cenko:2011:aoo}, etc. In particular, most of these processes can be simulated numerically, which provides ample physical information that can help us to choose appropriate kernels, either stationary or non-stationary, smooth or continuous, periodic or chirping, etc \cite{rasmussen:2006:gpf}, and makes Bayesian nonparametrics a promising data analysis methodology in physics and astronomy.



\begin{acknowledgments}
I thank my advisor Prof. Lee Samuel Finn for fruitful discussions on Bayesian nonparametrics analysis and valuable suggestions on the manuscript. I also thank Prof. Michael Eracleous for insightful comments on treating periapsis passages of highly eccentric binaries as gravitational wave burst sources, and Prof. Duncan Fong for suggestions on using Deviance Information Criterion as the criterion for Bayesian model comparison. This work was supported by Research Assistantship in the department of physics, and National Science Foundation Grant Numbers 09-40924 and 09-69857 awarded to The Pennsylvania State University.
\end{acknowledgments}


\begin{thebibliography}{45}%
\makeatletter
\providecommand \@ifxundefined [1]{%
 \@ifx{#1\undefined}
}%
\providecommand \@ifnum [1]{%
 \ifnum #1\expandafter \@firstoftwo
 \else \expandafter \@secondoftwo
 \fi
}%
\providecommand \@ifx [1]{%
 \ifx #1\expandafter \@firstoftwo
 \else \expandafter \@secondoftwo
 \fi
}%
\providecommand \natexlab [1]{#1}%
\providecommand \enquote  [1]{``#1''}%
\providecommand \bibnamefont  [1]{#1}%
\providecommand \bibfnamefont [1]{#1}%
\providecommand \citenamefont [1]{#1}%
\providecommand \href@noop [0]{\@secondoftwo}%
\providecommand \href [0]{\begingroup \@sanitize@url \@href}%
\providecommand \@href[1]{\@@startlink{#1}\@@href}%
\providecommand \@@href[1]{\endgroup#1\@@endlink}%
\providecommand \@sanitize@url [0]{\catcode `\\12\catcode `\$12\catcode
  `\&12\catcode `\#12\catcode `\^12\catcode `\_12\catcode `\%12\relax}%
\providecommand \@@startlink[1]{}%
\providecommand \@@endlink[0]{}%
\providecommand \url  [0]{\begingroup\@sanitize@url \@url }%
\providecommand \@url [1]{\endgroup\@href {#1}{\urlprefix }}%
\providecommand \urlprefix  [0]{URL }%
\providecommand \Eprint [0]{\href }%
\providecommand \doibase [0]{http://dx.doi.org/}%
\providecommand \selectlanguage [0]{\@gobble}%
\providecommand \bibinfo  [0]{\@secondoftwo}%
\providecommand \bibfield  [0]{\@secondoftwo}%
\providecommand \translation [1]{[#1]}%
\providecommand \BibitemOpen [0]{}%
\providecommand \bibitemStop [0]{}%
\providecommand \bibitemNoStop [0]{.\EOS\space}%
\providecommand \EOS [0]{\spacefactor3000\relax}%
\providecommand \BibitemShut  [1]{\csname bibitem#1\endcsname}%
\let\auto@bib@innerbib\@empty
\bibitem [{\citenamefont {Seto}()}]{Seto:2009:sfm}%
  \BibitemOpen
  \bibfield  {author} {\bibinfo {author} {\bibfnamefont {N.}~\bibnamefont
  {Seto}},\ }\href@noop {} {\bibfield  {journal} {\bibinfo  {journal}
  {Mon.~Not.~R.~Astron.~Soc.}\ }\textbf {\bibinfo {volume} {400}},\ \bibinfo
  {pages} {L38}}\BibitemShut {NoStop}%
\bibitem [{\citenamefont {Pshirkov}\ \emph {et~al.}(2010)\citenamefont
  {Pshirkov}, \citenamefont {Baskaran},\ and\ \citenamefont
  {Postnov}}]{Pshirkov:2010:ogw}%
  \BibitemOpen
  \bibfield  {author} {\bibinfo {author} {\bibfnamefont {M.~S.}\ \bibnamefont
  {Pshirkov}}, \bibinfo {author} {\bibfnamefont {D.}~\bibnamefont {Baskaran}},
  \ and\ \bibinfo {author} {\bibfnamefont {K.~A.}\ \bibnamefont {Postnov}},\
  }\href@noop {} {\bibfield  {journal} {\bibinfo  {journal}
  {Mon.~Not.~R.~Astron.~Soc.}\ }\textbf {\bibinfo {volume} {402}},\ \bibinfo
  {pages} {417} (\bibinfo {year} {2010})}\BibitemShut {NoStop}%
\bibitem [{\citenamefont {Cordes}\ and\ \citenamefont
  {Jenet}(2012)}]{cordes:2012:dgw}%
  \BibitemOpen
  \bibfield  {author} {\bibinfo {author} {\bibfnamefont {J.~M.}\ \bibnamefont
  {Cordes}}\ and\ \bibinfo {author} {\bibfnamefont {F.~A.}\ \bibnamefont
  {Jenet}},\ }\href@noop {} {\bibfield  {journal} {\bibinfo  {journal}
  {Astrophys. J.}\ }\textbf {\bibinfo {volume} {752}},\ \bibinfo {pages} {54}
  (\bibinfo {year} {2012})}\BibitemShut {NoStop}%
\bibitem [{\citenamefont {Damour}\ and\ \citenamefont
  {Vilenkin}(2001)}]{damour:2001:gwb}%
  \BibitemOpen
  \bibfield  {author} {\bibinfo {author} {\bibfnamefont {T.}~\bibnamefont
  {Damour}}\ and\ \bibinfo {author} {\bibfnamefont {A.}~\bibnamefont
  {Vilenkin}},\ }\href@noop {} {\bibfield  {journal} {\bibinfo  {journal}
  {Phys. Rev. D}\ }\textbf {\bibinfo {volume} {64}},\ \bibinfo {pages} {064008}
  (\bibinfo {year} {2001})}\BibitemShut {NoStop}%
\bibitem [{\citenamefont {Siemens}\ \emph {et~al.}(2007)\citenamefont
  {Siemens}, \citenamefont {Mandic},\ and\ \citenamefont
  {Creighton}}]{siemens:2007:gws}%
  \BibitemOpen
  \bibfield  {author} {\bibinfo {author} {\bibfnamefont {X.}~\bibnamefont
  {Siemens}}, \bibinfo {author} {\bibfnamefont {V.}~\bibnamefont {Mandic}}, \
  and\ \bibinfo {author} {\bibfnamefont {J.}~\bibnamefont {Creighton}},\
  }\href@noop {} {\bibfield  {journal} {\bibinfo  {journal} {Phys. Rev. Lett.}\
  }\textbf {\bibinfo {volume} {98}},\ \bibinfo {pages} {111101} (\bibinfo
  {year} {2007})}\BibitemShut {NoStop}%
\bibitem [{\citenamefont {Leblond}\ \emph {et~al.}(2009)\citenamefont
  {Leblond}, \citenamefont {Shlaer},\ and\ \citenamefont
  {Siemens}}]{leblond:2009:gwf}%
  \BibitemOpen
  \bibfield  {author} {\bibinfo {author} {\bibfnamefont {L.}~\bibnamefont
  {Leblond}}, \bibinfo {author} {\bibfnamefont {B.}~\bibnamefont {Shlaer}}, \
  and\ \bibinfo {author} {\bibfnamefont {X.}~\bibnamefont {Siemens}},\
  }\href@noop {} {\bibfield  {journal} {\bibinfo  {journal} {Phys. Rev. D}\ }\textbf
  {\bibinfo {volume} {79}},\ \bibinfo {pages} {123519} (\bibinfo {year}
  {2009})}\BibitemShut {NoStop}%
\bibitem [{\citenamefont {Finn}\ and\ \citenamefont
  {Lommen}(2010)}]{finn:2010:dla}%
  \BibitemOpen
  \bibfield  {author} {\bibinfo {author} {\bibfnamefont {L.~S.}\ \bibnamefont
  {Finn}}\ and\ \bibinfo {author} {\bibfnamefont {A.~N.}\ \bibnamefont
  {Lommen}},\ }\href@noop {} {\bibfield  {journal} {\bibinfo  {journal}
  {Astrophys. J.}\ }\textbf {\bibinfo {volume} {718}},\ \bibinfo {pages} {1400}
  (\bibinfo {year} {2010})}\BibitemShut {NoStop}%
\bibitem [{\citenamefont {Gelman}\ \emph {et~al.}(2004)\citenamefont {Gelman},
  \citenamefont {Carlin}, \citenamefont {Stern},\ and\ \citenamefont
  {Rubin}}]{gelman:2004:bda}%
  \BibitemOpen
  \bibfield  {author} {\bibinfo {author} {\bibfnamefont {A.}~\bibnamefont
  {Gelman}}, \bibinfo {author} {\bibfnamefont {J.~B.}\ \bibnamefont {Carlin}},
  \bibinfo {author} {\bibfnamefont {H.~S.}\ \bibnamefont {Stern}}, \ and\
  \bibinfo {author} {\bibfnamefont {D.~B.}\ \bibnamefont {Rubin}},\ }\href@noop
  {} {\emph {\bibinfo {title} {Bayesian Data Analysis}}},\ \bibinfo {edition}
  {2nd}\ ed.,\ Texts in Statistical Science\ (\bibinfo  {publisher} {Chapman \&
  Hall/CRC},\ \bibinfo {address} {Boca Raton, FL},\ \bibinfo {year}
  {2004})\BibitemShut {NoStop}%
\bibitem [{\citenamefont {Rasmussen}\ and\ \citenamefont
  {Christopher}(2006)}]{rasmussen:2006:gpf}%
  \BibitemOpen
  \bibfield  {author} {\bibinfo {author} {\bibfnamefont {C.~E.}\ \bibnamefont
  {Rasmussen}}\ and\ \bibinfo {author} {\bibfnamefont {K.~I.~W.}\ \bibnamefont
  {Christopher}},\ }\href@noop {} {\emph {\bibinfo {title} {Gaussian Process
  for Machine Learning}}},\ Adaptive Computation and Machine Learning\
  (\bibinfo  {publisher} {The MIT Press},\ \bibinfo {address} {Cambridge, MA},\
  \bibinfo {year} {2006})\BibitemShut {NoStop}%
\bibitem [{\citenamefont {Spiegelhalter}\ \emph {et~al.}(2002)\citenamefont
  {Spiegelhalter}, \citenamefont {Best}, \citenamefont {Carlin},\ and\
  \citenamefont {Linde}}]{spiegelhalter:2002:bmo}%
  \BibitemOpen
  \bibfield  {author} {\bibinfo {author} {\bibfnamefont {D.~J.}\ \bibnamefont
  {Spiegelhalter}}, \bibinfo {author} {\bibfnamefont {N.~G.}\ \bibnamefont
  {Best}}, \bibinfo {author} {\bibfnamefont {B.~P.}\ \bibnamefont {Carlin}}, \
  and\ \bibinfo {author} {\bibfnamefont {A.~V.~D.}\ \bibnamefont {Linde}},\
  }\href@noop {} {\bibfield  {journal} {\bibinfo  {journal}
  {J.~R.~Stat.~Soc.~B.,~Part~4}\ }\textbf {\bibinfo {volume} {64}},\ \bibinfo
  {pages} {583} (\bibinfo {year} {2002})}\BibitemShut {NoStop}%
\bibitem [{\citenamefont {Gibbons}\ and\ \citenamefont
  {Chakraborti}(2003)}]{gibbons:2003:nsi}%
  \BibitemOpen
  \bibfield  {author} {\bibinfo {author} {\bibfnamefont {J.~D.}\ \bibnamefont
  {Gibbons}}\ and\ \bibinfo {author} {\bibfnamefont {S.}~\bibnamefont
  {Chakraborti}},\ }\href@noop {} {\emph {\bibinfo {title} {Nonparametric
  Statistical Inference}}},\ \bibinfo {edition} {4th}\ ed.\ (\bibinfo
  {publisher} {Chapman \& Hall/CRC},\ \bibinfo {address} {Boca Raton, FL},\
  \bibinfo {year} {2003})\BibitemShut {NoStop}%
\bibitem [{\citenamefont {Ferguson}(1973)}]{ferguson:1973:aba}%
  \BibitemOpen
  \bibfield  {author} {\bibinfo {author} {\bibfnamefont {T.~S.}\ \bibnamefont
  {Ferguson}},\ }\href@noop {} {\bibfield  {journal} {\bibinfo  {journal} {The
  Annals of Statistics}\ }\textbf {\bibinfo {volume} {1}},\ \bibinfo {pages}
  {209} (\bibinfo {year} {1973})}\BibitemShut {NoStop}%
\bibitem [{\citenamefont {Doksum}(1974)}]{doksum:1974:tan}%
  \BibitemOpen
  \bibfield  {author} {\bibinfo {author} {\bibfnamefont {K.~A.}\ \bibnamefont
  {Doksum}},\ }\href@noop {} {\bibfield  {journal} {\bibinfo  {journal} {The
  Annals of Statistics}\ }\textbf {\bibinfo {volume} {2}},\ \bibinfo {pages}
  {183} (\bibinfo {year} {1974})}\BibitemShut {NoStop}%
\bibitem [{\citenamefont {O'Hagan}(1978)}]{o'hagan:1978:cfa}%
  \BibitemOpen
  \bibfield  {author} {\bibinfo {author} {\bibfnamefont {A.}~\bibnamefont
  {O'Hagan}},\ }\href@noop {} {\bibfield  {journal} {\bibinfo  {journal}
  {J.~R.~Stat.~Soc.~B}\ }\textbf {\bibinfo {volume} {40}},\ \bibinfo {pages}
  {1} (\bibinfo {year} {1978})}\BibitemShut {NoStop}%
\bibitem [{\citenamefont {Hjort}\ \emph {et~al.}(2010)\citenamefont {Hjort},
  \citenamefont {Holmes}, \citenamefont {M{\"u}ller},\ and\ \citenamefont
  {Walker}}]{hjort:2010:bn}%
  \BibitemOpen
  \bibinfo {editor} {\bibfnamefont {N.~L.}\ \bibnamefont {Hjort}}, \bibinfo
  {editor} {\bibfnamefont {C.}~\bibnamefont {Holmes}}, \bibinfo {editor}
  {\bibfnamefont {P.}~\bibnamefont {M{\"u}ller}}, \ and\ \bibinfo {editor}
  {\bibfnamefont {S.~G.}\ \bibnamefont {Walker}},\ eds.,\ \href@noop {} {\emph
  {\bibinfo {title} {Bayesian Nonparametrics}}},\ Cambridge Series in
  Statistical and Probabilistic Mathematics\ (\bibinfo  {publisher} {Cambridge
  University Press},\ \bibinfo {address} {Cambridge, UK},\ \bibinfo {year}
  {2010})\BibitemShut {NoStop}%
\bibitem [{\citenamefont {Ghosh}\ and\ \citenamefont
  {Ramamoorthi}(2003)}]{ghosh:2003:bn}%
  \BibitemOpen
  \bibfield  {author} {\bibinfo {author} {\bibfnamefont {J.~K.}\ \bibnamefont
  {Ghosh}}\ and\ \bibinfo {author} {\bibfnamefont {R.~V.}\ \bibnamefont
  {Ramamoorthi}},\ }\href@noop {} {\emph {\bibinfo {title} {Bayesian
  Nonparametrics}}},\ Springer Series in Statistics\ (\bibinfo  {publisher}
  {Springer-Verlag},\ \bibinfo {address} {New York, NY},\ \bibinfo {year}
  {2003})\BibitemShut {NoStop}%
\bibitem [{\citenamefont {Sudderth}(2006)}]{sudderth:2006:gmf}%
  \BibitemOpen
  \bibfield  {author} {\bibinfo {author} {\bibfnamefont {E.~B.}\ \bibnamefont
  {Sudderth}},\ }\emph {\bibinfo {title} {Graphical Models for Visual Object
  Recognition and Tracking}},\ \href@noop {} {Ph.D. thesis},\ \bibinfo
  {school} {Masschusetts Institute of Technology}, \bibinfo {address}
  {Cambridge, MA} (\bibinfo {year} {2006})\BibitemShut {NoStop}%
\bibitem [{\citenamefont {Summerscales}\ \emph {et~al.}(2008)\citenamefont
  {Summerscales}, \citenamefont {Burrows}, \citenamefont {Finn},\ and\
  \citenamefont {Ott}}]{summerscales:2008:mef}%
  \BibitemOpen
  \bibfield  {author} {\bibinfo {author} {\bibfnamefont {T.~Z.}\ \bibnamefont
  {Summerscales}}, \bibinfo {author} {\bibfnamefont {A.}~\bibnamefont
  {Burrows}}, \bibinfo {author} {\bibfnamefont {L.~S.}\ \bibnamefont {Finn}}, \
  and\ \bibinfo {author} {\bibfnamefont {C.~D.}\ \bibnamefont {Ott}},\
  }\href@noop {} {\bibfield  {journal} {\bibinfo  {journal} {Astrophys. J.}\
  }\textbf {\bibinfo {volume} {678}},\ \bibinfo {pages} {1142} (\bibinfo {year}
  {2008})}\BibitemShut {NoStop}%
\bibitem [{\citenamefont {Bretthorst}(1988)}]{Bretthorst:1988:bsa}%
  \BibitemOpen
  \bibfield  {author} {\bibinfo {author} {\bibfnamefont {G.~L.}\ \bibnamefont
  {Bretthorst}},\ }\href@noop {} {\emph {\bibinfo {title} {Bayesian Spectrum
  Analysis and Parameter Estimation}}},\ Springer Series in Statistics\
  (\bibinfo  {publisher} {Springer},\ \bibinfo {address} {New York, NY},\
  \bibinfo {year} {1988})\BibitemShut {NoStop}%
\bibitem [{\citenamefont {Blanchet}(2006)}]{blanchet:2006:grf}%
  \BibitemOpen
  \bibfield  {author} {\bibinfo {author} {\bibfnamefont {L.}~\bibnamefont
  {Blanchet}},\ }\href@noop {} {\bibfield  {journal} {\bibinfo  {journal}
  {Living.~Rev.~Relativity}\ }\textbf {\bibinfo {volume} {9}},\ \bibinfo
  {pages} {4} (\bibinfo {year} {2006})}\BibitemShut {NoStop}%
\bibitem [{\citenamefont {Misner}\ \emph {et~al.}(1973)\citenamefont {Misner},
  \citenamefont {Thorne},\ and\ \citenamefont {Wheeler}}]{misner:1973:g}%
  \BibitemOpen
  \bibfield  {author} {\bibinfo {author} {\bibfnamefont {C.~W.}\ \bibnamefont
  {Misner}}, \bibinfo {author} {\bibfnamefont {K.~S.}\ \bibnamefont {Thorne}},
  \ and\ \bibinfo {author} {\bibfnamefont {J.~A.}\ \bibnamefont {Wheeler}},\
  }\href@noop {} {\emph {\bibinfo {title} {Gravitation}}}\ (\bibinfo
  {publisher} {W.~H.~Freeman and Company},\ \bibinfo {address} {New York, NY},\
  \bibinfo {year} {1973})\BibitemShut {NoStop}%
\bibitem [{\citenamefont {Pitkin}(2012)}]{pitkin:2012:egw}%
  \BibitemOpen
  \bibfield  {author} {\bibinfo {author} {\bibfnamefont {M.}~\bibnamefont
  {Pitkin}},\ }\href@noop {} {\bibfield  {journal} {\bibinfo  {journal}
  {Mon.~Not.~R.~Astron.~Soc.}\ }\textbf {\bibinfo {volume} {425}},\ \bibinfo
  {pages} {2688} (\bibinfo {year} {2012})}\BibitemShut {NoStop}%
\bibitem [{\citenamefont {Hobbs}\ \emph {et~al.}(2010)\citenamefont {Hobbs}
  \emph {et~al.}}]{hobbs:2010:tip}%
  \BibitemOpen
  \bibfield  {author} {\bibinfo {author} {\bibfnamefont {G.~B.}\ \bibnamefont
  {Hobbs}} \emph {et~al.},\ }\href@noop {} {\bibfield  {journal} {\bibinfo
  {journal} {Class. Quant. Grav.}\ }\textbf {\bibinfo {volume} {27}},\ \bibinfo
  {pages} {084043} (\bibinfo {year} {2010})}\BibitemShut {NoStop}%
\bibitem [{\citenamefont {Demorest}\ \emph {et~al.}(2012)\citenamefont
  {Demorest} \emph {et~al.}}]{demorest:2012:lot}%
  \BibitemOpen
  \bibfield  {author} {\bibinfo {author} {\bibfnamefont {P.~B.}\ \bibnamefont
  {Demorest}} \emph {et~al.},\ }\href@noop {} {\bibfield  {journal} {\bibinfo
  {journal} {Astrophys. J.}\ }\textbf {\bibinfo {volume} {762}},\ \bibinfo
  {pages} {94} (\bibinfo {year} {2012})}\BibitemShut {NoStop}%
\bibitem [{\citenamefont {Manchester}\ \emph {et~al.}()\citenamefont
  {Manchester} \emph {et~al.}}]{manchester:2012:tpp}%
  \BibitemOpen
  \bibfield  {author} {\bibinfo {author} {\bibfnamefont {R.~N.}\ \bibnamefont
  {Manchester}} \emph {et~al.},\ }\href@noop {} {}\bibinfo {note}
  {\lowercase{a}rXiv:~1210.6130, accepted by PASA}\BibitemShut {NoStop}%
\bibitem [{\citenamefont {Adler}(1981)}]{adler:1981:tgo}%
  \BibitemOpen
  \bibfield  {author} {\bibinfo {author} {\bibfnamefont {R.~J.}\ \bibnamefont
  {Adler}},\ }\href@noop {} {\emph {\bibinfo {title} {The Geometry of Random
  Fields}}},\ Wiley Series in Probability and Mathematical Statistics\
  (\bibinfo  {publisher} {John Wiley \& Sons},\ \bibinfo {address} {Chichester,
  UK},\ \bibinfo {year} {1981})\BibitemShut {NoStop}%
\bibitem [{\citenamefont {Stein}(1999)}]{stein:1999:ios}%
  \BibitemOpen
  \bibfield  {author} {\bibinfo {author} {\bibfnamefont {M.~L.}\ \bibnamefont
  {Stein}},\ }\href@noop {} {\emph {\bibinfo {title} {Interpolation of Spatial
  Data}}},\ Springer Series in Statistics\ (\bibinfo  {publisher} {Springer},\
  \bibinfo {address} {New York, NY},\ \bibinfo {year} {1999})\BibitemShut
  {NoStop}%
\bibitem [{\citenamefont {Gelman}(2006)}]{gelman:2006:pdf}%
  \BibitemOpen
  \bibfield  {author} {\bibinfo {author} {\bibfnamefont {A.}~\bibnamefont
  {Gelman}},\ }\href@noop {} {\bibfield  {journal} {\bibinfo  {journal}
  {Bayesian.~Anal.}\ }\textbf {\bibinfo {volume} {1}},\ \bibinfo {pages} {515}
  (\bibinfo {year} {2006})}\BibitemShut {NoStop}%
\bibitem [{\citenamefont {Jeffreys}(1946)}]{jeffreys:1946:aif}%
  \BibitemOpen
  \bibfield  {author} {\bibinfo {author} {\bibfnamefont {H.}~\bibnamefont
  {Jeffreys}},\ }\href@noop {} {\bibfield  {journal} {\bibinfo  {journal}
  {Proc.~R.~Soc.~London.,~Ser.~A}\ }\textbf {\bibinfo {volume} {186}},\
  \bibinfo {pages} {453} (\bibinfo {year} {1946})}\BibitemShut {NoStop}%
\bibitem [{\citenamefont {Kass}\ and\ \citenamefont
  {Raftery}(1995)}]{kass:1995:bf}%
  \BibitemOpen
  \bibfield  {author} {\bibinfo {author} {\bibfnamefont {R.~E.}\ \bibnamefont
  {Kass}}\ and\ \bibinfo {author} {\bibfnamefont {A.~E.}\ \bibnamefont
  {Raftery}},\ }\href@noop {} {\bibfield  {journal} {\bibinfo  {journal}
  {J.~Am.~Stat.~Assoc.}\ }\textbf {\bibinfo {volume} {90}},\ \bibinfo {pages}
  {773} (\bibinfo {year} {1995})}\BibitemShut {NoStop}%
\bibitem [{\citenamefont {Kullback}\ and\ \citenamefont
  {Leibler}(1951)}]{kullback:1951:oia}%
  \BibitemOpen
  \bibfield  {author} {\bibinfo {author} {\bibfnamefont {S.}~\bibnamefont
  {Kullback}}\ and\ \bibinfo {author} {\bibfnamefont {R.~A.}\ \bibnamefont
  {Leibler}},\ }\href@noop {} {\bibfield  {journal} {\bibinfo  {journal}
  {Annu.~Math.~Stat.}\ }\textbf {\bibinfo {volume} {22}},\ \bibinfo {pages}
  {79} (\bibinfo {year} {1951})}\BibitemShut {NoStop}%
\bibitem [{\citenamefont {Claeskens}\ and\ \citenamefont
  {Hjort}(2008)}]{claeskens:2008:msa}%
  \BibitemOpen
  \bibfield  {author} {\bibinfo {author} {\bibfnamefont {G.}~\bibnamefont
  {Claeskens}}\ and\ \bibinfo {author} {\bibfnamefont {N.~L.}\ \bibnamefont
  {Hjort}},\ }\href@noop {} {\emph {\bibinfo {title} {Model Selection and Model
  Averaging}}},\ Cambridge Series in Statistical and Probabilistic Mathematics\
  (\bibinfo  {publisher} {Cambridge University Press},\ \bibinfo {address}
  {Cambridge, UK},\ \bibinfo {year} {2008})\BibitemShut {NoStop}%
\bibitem [{\citenamefont {Jefferys}\ and\ \citenamefont
  {Berger}(1991)}]{jefferys:1991:sor}%
  \BibitemOpen
  \bibfield  {author} {\bibinfo {author} {\bibfnamefont {W.~H.}\ \bibnamefont
  {Jefferys}}\ and\ \bibinfo {author} {\bibfnamefont {J.~O.}\ \bibnamefont
  {Berger}},\ }\href@noop {} {\emph {\bibinfo {title} {Sharpening Ockham's
  Razor on a Bayesian Strop}}},\ \bibinfo {type} {Tech. Rep.}\ \bibinfo
  {number} {91-44C}\ (\bibinfo  {institution} {Department of Statistics, Purdue
  University},\ \bibinfo {year} {1991})\BibitemShut {NoStop}%
\bibitem [{\citenamefont {Meng}\ and\ \citenamefont
  {Rubin}(1992)}]{meng:1992:plr}%
  \BibitemOpen
  \bibfield  {author} {\bibinfo {author} {\bibfnamefont {X.}~\bibnamefont
  {Meng}}\ and\ \bibinfo {author} {\bibfnamefont {D.~B.}\ \bibnamefont
  {Rubin}},\ }\href@noop {} {\bibfield  {journal} {\bibinfo  {journal}
  {Biometrika}\ }\textbf {\bibinfo {volume} {79}},\ \bibinfo {pages} {103}
  (\bibinfo {year} {1992})}\BibitemShut {NoStop}%
\bibitem [{\citenamefont {Burt}\ \emph {et~al.}(2011)\citenamefont {Burt},
  \citenamefont {Lommen},\ and\ \citenamefont {Finn}}]{burt:2011:opt}%
  \BibitemOpen
  \bibfield  {author} {\bibinfo {author} {\bibfnamefont {B.~J.}\ \bibnamefont
  {Burt}}, \bibinfo {author} {\bibfnamefont {A.~N.}\ \bibnamefont {Lommen}}, \
  and\ \bibinfo {author} {\bibfnamefont {L.~S.}\ \bibnamefont {Finn}},\
  }\href@noop {} {\bibfield  {journal} {\bibinfo  {journal} {Astrophys. J.}\
  }\textbf {\bibinfo {volume} {730}},\ \bibinfo {pages} {17} (\bibinfo {year}
  {2011})}\BibitemShut {NoStop}%
\bibitem [{\citenamefont {Hobbs}\ \emph {et~al.}(2006)\citenamefont {Hobbs},
  \citenamefont {Edwards},\ and\ \citenamefont {Manchester}}]{hobbs:2006:tan}%
  \BibitemOpen
  \bibfield  {author} {\bibinfo {author} {\bibfnamefont {G.~B.}\ \bibnamefont
  {Hobbs}}, \bibinfo {author} {\bibfnamefont {R.~T.}\ \bibnamefont {Edwards}},
  \ and\ \bibinfo {author} {\bibfnamefont {R.~T.}\ \bibnamefont {Manchester}},\
  }\href@noop {} {\bibfield  {journal} {\bibinfo  {journal}
  {Mon.~Not.~R.~Astron.~Soc.}\ }\textbf {\bibinfo {volume} {369}},\ \bibinfo
  {pages} {655} (\bibinfo {year} {2006})}\BibitemShut {NoStop}%
\bibitem [{\citenamefont {Edwards}\ \emph {et~al.}(2006)\citenamefont
  {Edwards}, \citenamefont {Hobbs},\ and\ \citenamefont
  {Manchester}}]{edwards:2006:tan}%
  \BibitemOpen
  \bibfield  {author} {\bibinfo {author} {\bibfnamefont {R.~T.}\ \bibnamefont
  {Edwards}}, \bibinfo {author} {\bibfnamefont {G.~B.}\ \bibnamefont {Hobbs}},
  \ and\ \bibinfo {author} {\bibfnamefont {R.~T.}\ \bibnamefont {Manchester}},\
  }\href@noop {} {\bibfield  {journal} {\bibinfo  {journal}
  {Mon.~Not.~R.~Astron.~Soc.}\ }\textbf {\bibinfo {volume} {369}},\ \bibinfo
  {pages} {655} (\bibinfo {year} {2006})}\BibitemShut {NoStop}%
\bibitem [{\citenamefont {Coles}\ \emph {et~al.}(2011)\citenamefont {Coles},
  \citenamefont {Hobbs}, \citenamefont {Champion}, \citenamefont {Manchester},\
  and\ \citenamefont {Verbiest}}]{coles:2011:pta}%
  \BibitemOpen
  \bibfield  {author} {\bibinfo {author} {\bibfnamefont {W.}~\bibnamefont
  {Coles}}, \bibinfo {author} {\bibfnamefont {G.}~\bibnamefont {Hobbs}},
  \bibinfo {author} {\bibfnamefont {D.~J.}\ \bibnamefont {Champion}}, \bibinfo
  {author} {\bibfnamefont {R.}~\bibnamefont {Manchester}}, \ and\ \bibinfo
  {author} {\bibfnamefont {J.~P.~W.}\ \bibnamefont {Verbiest}},\ }\href@noop {}
  {\bibfield  {journal} {\bibinfo  {journal} {Mon.~Not.~R.~Astron.~Soc.}\
  }\textbf {\bibinfo {volume} {418}},\ \bibinfo {pages} {561} (\bibinfo {year}
  {2011})}\BibitemShut {NoStop}%
\bibitem [{\citenamefont {Shannon}\ and\ \citenamefont
  {Cordes}(2010)}]{shannon:2010:atr}%
  \BibitemOpen
  \bibfield  {author} {\bibinfo {author} {\bibfnamefont {R.~M.}\ \bibnamefont
  {Shannon}}\ and\ \bibinfo {author} {\bibfnamefont {J.~M.}\ \bibnamefont
  {Cordes}},\ }\href@noop {} {\bibfield  {journal} {\bibinfo  {journal}
  {Astrophys. J.}\ }\textbf {\bibinfo {volume} {725}},\ \bibinfo {pages} {1607}
  (\bibinfo {year} {2010})}\BibitemShut {NoStop}%
\bibitem [{\citenamefont {Robert}\ and\ \citenamefont
  {Casella}()}]{robert:2004:mcs}%
  \BibitemOpen
  \bibfield  {author} {\bibinfo {author} {\bibfnamefont {C.~P.}\ \bibnamefont
  {Robert}}\ and\ \bibinfo {author} {\bibfnamefont {G.}~\bibnamefont
  {Casella}},\ }\href@noop {} {\emph {\bibinfo {title} {Monte Carlo Statistical
  Methods}}},\ \bibinfo {edition} {2nd}\ ed.,\ Springer Series in Statistics\
  (\bibinfo  {publisher} {Springer},\ \bibinfo {address} {New York,
  NY})\BibitemShut {NoStop}%
\bibitem [{\citenamefont {Nan}\ \emph {et~al.}(2011)\citenamefont {Nan},
  \citenamefont {Li}, \citenamefont {Jin}, \citenamefont {Wang}, \citenamefont
  {Zhu}, \citenamefont {Zhu}, \citenamefont {Zhang}, \citenamefont {Yue},\ and\
  \citenamefont {Qian}}]{nan:2011:tfa}%
  \BibitemOpen
  \bibfield  {author} {\bibinfo {author} {\bibfnamefont {R.}~\bibnamefont
  {Nan}}, \bibinfo {author} {\bibfnamefont {D.}~\bibnamefont {Li}}, \bibinfo
  {author} {\bibfnamefont {C.}~\bibnamefont {Jin}}, \bibinfo {author}
  {\bibfnamefont {Q.}~\bibnamefont {Wang}}, \bibinfo {author} {\bibfnamefont
  {L.}~\bibnamefont {Zhu}}, \bibinfo {author} {\bibfnamefont {W.}~\bibnamefont
  {Zhu}}, \bibinfo {author} {\bibfnamefont {H.}~\bibnamefont {Zhang}}, \bibinfo
  {author} {\bibfnamefont {Y.}~\bibnamefont {Yue}}, \ and\ \bibinfo {author}
  {\bibfnamefont {L.}~\bibnamefont {Qian}},\ }\href@noop {} {\bibfield
  {journal} {\bibinfo  {journal} {Int.~J.~Mod.~Phys.~D}\ }\textbf {\bibinfo
  {volume} {20}},\ \bibinfo {pages} {989} (\bibinfo {year} {2011})}\BibitemShut
  {NoStop}%
\bibitem [{\citenamefont {Dewdney}\ \emph {et~al.}(2009)\citenamefont
  {Dewdney}, \citenamefont {Hall}, \citenamefont {Schilizzi},\ and\
  \citenamefont {Lazio}}]{dewdney:2009:tsk}%
  \BibitemOpen
  \bibfield  {author} {\bibinfo {author} {\bibfnamefont {P.~E.}\ \bibnamefont
  {Dewdney}}, \bibinfo {author} {\bibfnamefont {P.~J.}\ \bibnamefont {Hall}},
  \bibinfo {author} {\bibfnamefont {R.~T.}\ \bibnamefont {Schilizzi}}, \ and\
  \bibinfo {author} {\bibfnamefont {T.~J.~L.~W.}\ \bibnamefont {Lazio}},\
  }\href@noop {} {\bibfield  {journal} {\bibinfo  {journal} {Proc.~IEEE}\
  }\textbf {\bibinfo {volume} {97}},\ \bibinfo {pages} {1482} (\bibinfo {year}
  {2009})}\BibitemShut {NoStop}%
\bibitem [{\citenamefont {Aasi}\ \emph {et~al.}(2013)\citenamefont {Aasi} \emph
  {et~al.}}]{aasi:2013:sfg}%
  \BibitemOpen
  \bibfield  {author} {\bibinfo {author} {\bibfnamefont {J.}~\bibnamefont
  {Aasi}} \emph {et~al.},\ }\href@noop {} {\bibfield  {journal} {\bibinfo
  {journal} {Phys. Rev. D}\ }\textbf {\bibinfo {volume} {87}},\ \bibinfo
  {pages} {022002} (\bibinfo {year} {2013})}\BibitemShut {NoStop}%
\bibitem [{\citenamefont {Deck}\ \emph {et~al.}(2012)\citenamefont {Deck} \emph
  {et~al.}}]{deck:2012:rdc}%
  \BibitemOpen
  \bibfield  {author} {\bibinfo {author} {\bibfnamefont {K.~M.}\ \bibnamefont
  {Deck}} \emph {et~al.},\ }\href@noop {} {\bibfield  {journal} {\bibinfo
  {journal} {Astrophys. J.}\ }\textbf {\bibinfo {volume} {755}},\ \bibinfo
  {pages} {L21} (\bibinfo {year} {2012})}\BibitemShut {NoStop}%
\bibitem [{\citenamefont {Cenko}\ \emph {et~al.}(2011)\citenamefont {Cenko}
  \emph {et~al.}}]{cenko:2011:aoo}%
  \BibitemOpen
  \bibfield  {author} {\bibinfo {author} {\bibfnamefont {S.~B.}\ \bibnamefont
  {Cenko}} \emph {et~al.},\ }\href@noop {} {\bibfield  {journal} {\bibinfo
  {journal} {Astrophys. J.}\ }\textbf {\bibinfo {volume} {732}},\ \bibinfo
  {pages} {29} (\bibinfo {year} {2011})}\BibitemShut {NoStop}%
\end{thebibliography}
\end{document}